\documentclass[a4paper, 12pt]{article}
\usepackage[utf8]{inputenc}
\usepackage[english]{babel}
\usepackage{natbib}
\usepackage[margin=1in]{geometry}
\usepackage{graphicx}
\usepackage{subcaption}
\usepackage{rotating}
\usepackage{caption}
\usepackage[pagebackref=true]{hyperref}
\hypersetup{colorlinks=true, citecolor=blue}
\usepackage{setspace}
\graphicspath{ {Figures/} }

\title{The Legacy of Authoritarianism in a Democracy\thanks{I thank Rikhil Bhavnani, Charles Crabtree, Jun Goto, Masaaki Higashijima, Yusaku Horiuchi, Bhumi Purohit, Harunobu Saijo, Kimiko Terai,  Yasutora Watanabe, Adam Ziegfeld, and numerous seminar and conference participants for their excellent feedback. All remaining errors are my own. I am grateful for financial support from the Japan Society for the Promotion of Science (No. 21K13307). }}

\author{Pramod Kumar Sur\thanks{Ritsumeikan University and Osaka University Email: pramodsur@gmail.com}}

\date{June 2023}

\begin{document}
\maketitle

\begin{abstract}
\noindent
\linespread{1.0}\selectfont
Recent democratic backsliding and the rise of authoritarian regimes worldwide have rekindled interest in understanding the causes and consequences of such authoritarian rule in democracies. In this paper, I study the long-run political consequences of authoritarianism in the world's largest democracy. Exploiting the unexpected timing of the authoritarian rule imposed in India in the 1970s and using a difference-in-difference, triple difference, and a regression discontinuity design estimation approach, I document a sharp decline in the then-dominant incumbent, the Indian National Congress party's political dominance in subsequent years. I also present evidence that the decline in political dominance was not at the expense of a lower voter turnout rate. Instead, a sharp rise in the number of opposition candidates contesting elections in subsequent years played an important role. Finally, I examine the enduring consequences, revealing that confidence in politicians remains low in states where the draconian policy was high.
\end{abstract}
\textbf{Keywords:} Democracy, authoritarianism, voting behavior, voter turnout, confidence in institutions, family planning, India

\newpage

\section{Introduction}

\begin{figure}[htbp]
\begin{center}
\caption{\label{figure:Figure1}\textbf {Changes in Democracy Between 2012 and 2017}}
\includegraphics[width=\textwidth]{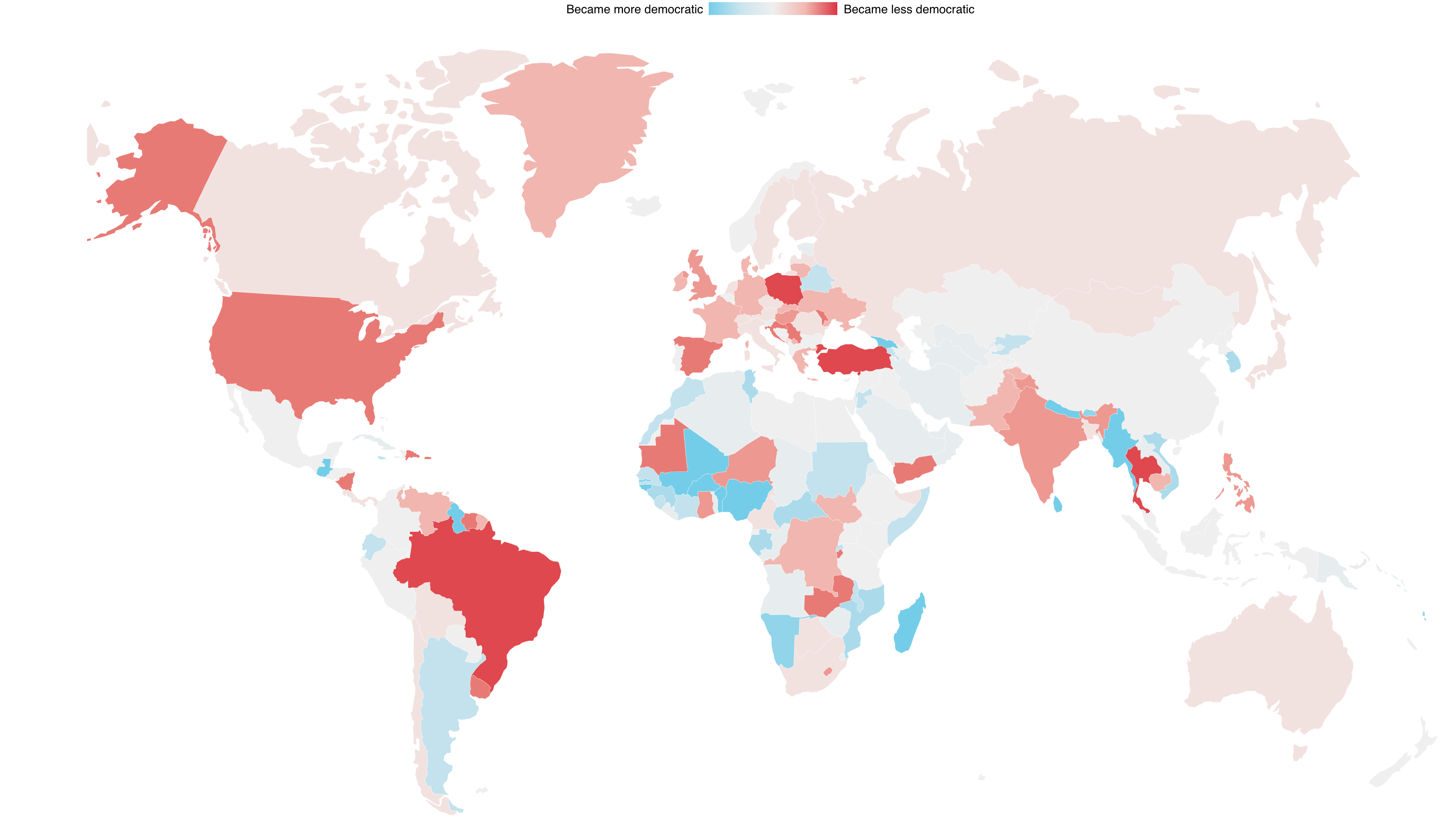}
\end{center}
{\small Notes: The figure presents the country-level change in the Liberal Democracy Index (LDI) between 2012 and 2017. Darker blue indicates countries that have become more democratic, and darker red indicates countries that have become less democratic. The LDI considers constitutionally protected civil liberties, a strong rule of law, an independent judiciary, and effective checks and balances that limit the exercise of executive power. The index also considers the level of electoral democracy. \\
\noindent Data Source: Varieties of Democracy (V-Dem).
https://www.v-dem.net/en/ \\
\noindent Figure Source:
https://www.bloomberg.com/graphics/2018-democracy-decline/ Accessed on 20th July 2021
}
\end{figure}

The world has been in the grip of a democratic recession and authoritarian regimes and autocratic rules are on the rise.\footnote{For example, see \citet{diamond_world_2021}, \citet{hodal_study_2019}, and \citet{hyde_democracys_2020}.}
Figure \ref{figure:Figure1} presents the change in the Liberal Democracy Index (LDI) of countries for the period 2012--2017 from the Varieties of Democracy database.\footnote{Figure \ref{figure:FigureA1} in the Appendix additionally presents the countries that have become substantially either autocratic or democratic between 2010 and 2020.}
Although several countries are transitioning toward more democratic rule (especially in Africa), the figure clearly suggests that even more countries have become less democratic over this period, especially in Asia, Europe, and the Americas.\footnote{For detail, see \citet{alizada_autocratization_2021}, \citet{bloomberg_elected_2018}, and \citet{luhrmann_third_2019}.}
Even more strikingly, recent democratic breakdowns and surges of authoritarian rule are emerging in well-established democracies (such as India) and in wealthier countries (such as Hungary, Poland, Turkey, and Venezuela).\footnote{Moreover, the recent responses to the COVID-19 pandemic have severely exacerbated the situation undermining democratic rule in many countries.} With more countries backsliding rather than improving in their overall democratic performance, and especially with mainstream political parties (and leaders) in these countries transitioning toward authoritarian rule, the global trend toward authoritarianism should be viewed with concern, and its consequences should be assessed appropriately.

\begin{center}
\centering Figure \ref{figure:Figure1} about here
\end{center}

\begin{figure}[htbp]
\begin{center}
\caption{\label{figure:Figure2}\textbf {Change in Democracy Index in India}}
\includegraphics[width=\textwidth]{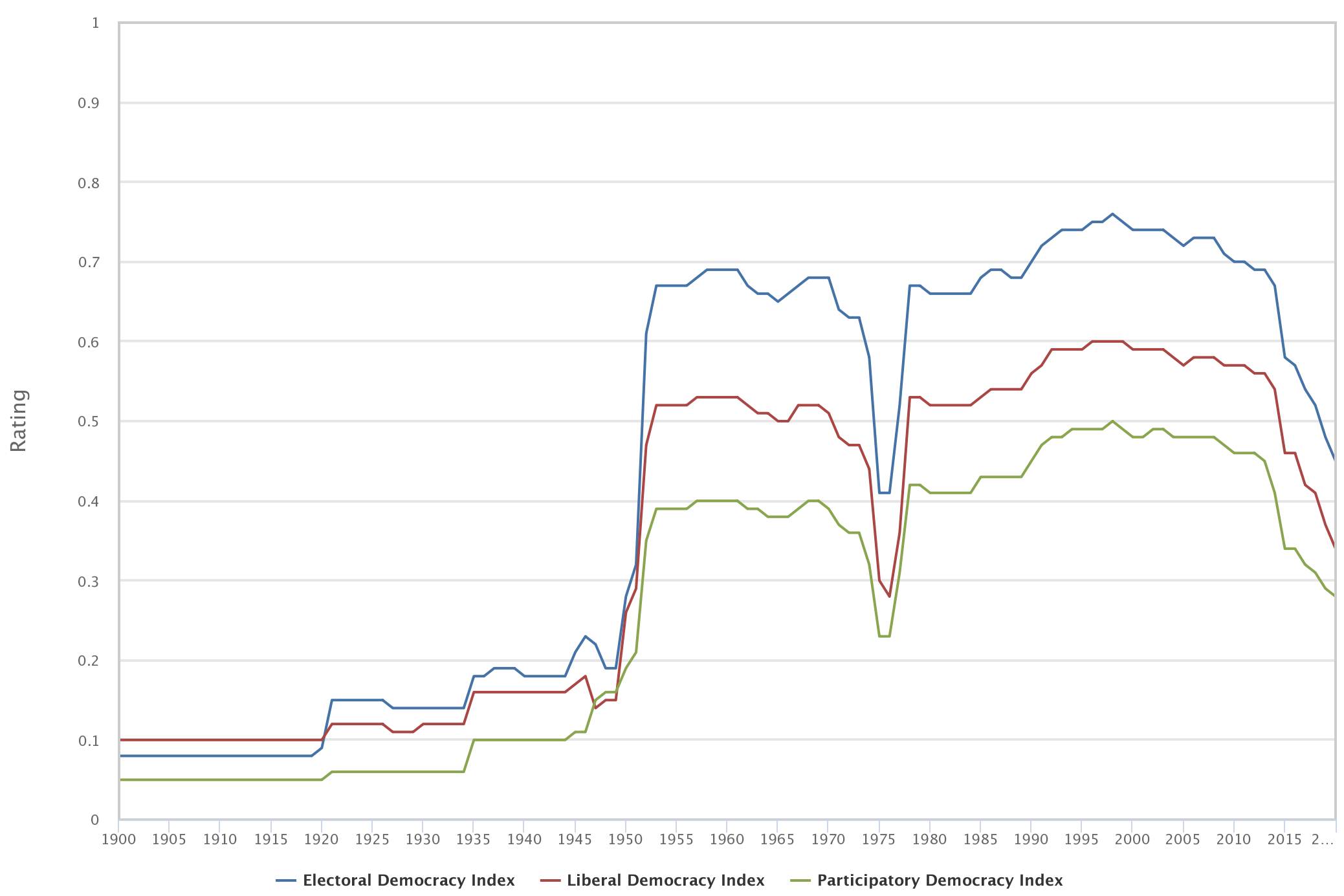}
\end{center}
{\small Notes: The figure presents the change in the Democracy Index in India between 1900 and 2020. The blue, red, and green lines indicate Electoral Democracy Index, LDI, and Participatory Democracy Index, respectively. \\
Data Source: Varieties of Democracy (V-Dem). https://www.v-dem.net/en/ Accessed on 20th July 2021
}
\end{figure}

In this article, I explore the political consequences of authoritarianism within a democracy in the context of India—the world's largest democracy. The unexpected authoritarian rule in India during the 1970s provides a unique natural setting in history to study the legacy of authoritarianism in a democracy. Between 1975 and 1977, India went through a brief period of authoritarian rule under Prime Minister Indira Gandhi of the ruling Indian National Congress (INC) party. Her draconian rule during this period was the first experience of electoral autocracy in India. On June 25, 1975, she unexpectedly proclaimed a state of emergency, popularly known as ``the emergency'' in India, which, under the Indian constitution, suspended a wide range of civil liberties. In the subsequent periods, she implemented several draconian policies, imposed a series of new legislation, and made constitutional amendments to govern the country. However, in January 1977, she unexpectedly called an election and lifted the emergency rule.\footnote{The emergency rule officially ended in March 1977 after the defeat of the INC in the lower house of the parliament.} Figure \ref{figure:Figure2} presents the (sudden) change in the democratic index during this brief period in India.

\begin{center}
\centering Figure \ref{figure:Figure2} about here
\end{center}

My empirical strategy is to investigate the impact of authoritarian rule on political outcomes in the subsequent periods in India. I interpret the authoritarian rule between 1975 and 1977 as an exogenous shock to Indian democracy and trace the political responses in subsequent periods. I use the universe of national and state-level election data between 1962 and 2023 to investigate the political consequences of authoritarian rule.

My main finding is that authoritarian rule has had a first-order impact on the decline of the political dominance of the then-incumbent party, the INC, in India. Using a difference-in-differences (DID) estimation approach, I document a significantly lower INC vote share and a lesser number of INC candidates winning elections in subsequent years. My estimates suggest that the authoritarian rule between 1975 and 1977 accounts for about a four-percentage point decline in the INC's vote share in the lower house of the Indian parliament (Lok Sabha) election in subsequent years. Furthermore, the draconian rule can explain as much as a 28-percentage point drop in the INC candidates' probability of winning elections in subsequent years.

I bolster this interpretation with several exercises. First, I show that an alternative source of variation of the authoritarian rule provides similar results. If the decline of the INC was partly a response to the brief period of authoritarianism, then a draconian policy implemented during this period intensifying the authoritarian rule should have similar results. I document that this is the case. I focus on an aggressive policy undertaken during this period, that is, the forced sterilization policy, and use the excess sterilizations performed at the state level as a second level of variation. As historians have argued, and as I will explain in detail in the next section, among all authoritarian policies, aggressive family planning through forced sterilization was a hallmark of the emergency period and a policy that directly affected the general population \citep{chandra_name_2017, dhar_indira_2018, gwatkin_political_1979, nayar_emergency_2013, panandiker_family_1978, shah_commission_of_inquiry_third_1978, williams_storming_2014}. Using this as my second-level variation in the authoritarian rule, I implement a triple-difference (DDD, henceforth) estimation strategy and find results consistent with my interpretation. Second, I estimate a flexible DDD model that allows for the estimation of election year-specific treatment effects through an event study design. I find that the estimates for pre-emergency periods, i.e., the years prior to 1977, fluctuate around zero and do not follow any specific trend. However, most post-emergency years' coefficients are negative and highly statistically significant. The legacy of the authoritarian rule persists and has continued to affect the INC's political representation up to the most recent election in 2019. In both cases, the results are consistent with a causal interpretation that the authoritarian rule imposed by the INC has had a first-order impact on the decline of its political dominance in India.

As a final step in my empirical estimation strategy, I turn to examine the impact of the authoritarian rule in India through a regression discontinuity design (RDD) approach using state-level election data (Assembly Elections) between 1962 and 2023. Similar to the national elections, I document a sharp discontinuity in the vote share and winning probability of INC candidates after the emergency rule. These results further bolster my interpretation that the authoritarian rule in India led to a sharp decline in the then-dominant incumbent, the Indian National Congress party's political dominance in subsequent years.

Then, I explore the plausible mechanisms. I first show that the decline in the INC's political dominance did not occur because of a lower voter turnout rate in subsequent elections. The voter turnout rate did not decline after the authoritarian rule, suggesting that voters switched their voting behavior favoring non-INC candidates. However, the autocratic rule gave rise to the formation of new political parties. In particular, I document a sharp increase in the number of candidates contesting elections from a constituency after 1977. This may have been because the authoritarian rule and draconian policy imposed during the emergency may have created a credibility gap in the INC's rule over India. As a result, it may have had a direct impact on the formation of new opposition parties as a direct deterrence to the INC rule, which thereby captured some of the INC votes.

Finally, I explore the durable consequences that endure in present-day India. Using state-level variation in the draconian forced sterilization policy, I document that households belonging to states heavily exposed to the draconian policy exhibit a lower level of confidence in politicians today. My interpretation is that the authoritarian rule and draconian policy implemented by the INC, then the largest political party in India, may have delegitimized the political establishment. Thus, the legacy of this historical event persists today.

This paper builds on and contributes to the political economy literature on authoritarian regimes in democracies. The recent democratic backsliding and the rise of electoral autocracies and authoritarian populist regimes has sparked a new generation of studies on the erosion of democracy and the process of autocratization \citep[e.g.,][]{bermeo_democratic_2016, coppedge_eroding_2017, diamond_democratic_2008, diamond_spirit_2008, diamond_facing_2015, hyde_democracys_2020, levitsky_how_2018, luhrmann_third_2019, mechkova_how_2017, runciman_how_2018-1}. In addition, they have rekindled our interest in understanding and exploring what lies behind the ascendancy of such regimes in democracies. Popular explanations include moral values, identity, and culture \citep[e.g.,][]{enke2020moral, margalit_economic_2019, mutz_status_2018, norris_cultural_2019, sides_identity_2019}, economic shocks such as financial crises \citep[e.g.,][]{algan_european_2017, dehdari_economic_2021, fetzer_did_2019, funke_going_2016, mian_resolving_2014}, trade and globalization \citep[e.g.,][]{autor_importing_2020, colantone_trade_2018}, automation \citep[e.g.,][]{anelli2019we, frey2018political, im2019losers}, immigration \citep[e.g.,][]{dustmann_refugee_2019, halla_immigration_2017, tabellini_gifts_2020}, social media and the internet \citep[e.g.,][]{zhuravskaya2020political}, and the interactions of this factors.\footnote {For a detailed review of literature in this field, particularly the recent rise of authoritarian populism, please refer to \citet{berman_causes_2021-1, guriev_political_2020, rodrik_why_2020}, and the references therein.} However, we have limited systematic evidence and few econometric studies about the political legacy of such regimes in the long run. Understanding the legacy is important as the institutional change caused by the authoritarian rule---both formal and informal---and citizens' experience with the regime may persist. Additionally, and more importantly, examining the \emph{long-run} legacy is relevant as institutions and social norms evolve slowly.  My paper is most closely related to the strand of political economy literature that focuses on the rise of fascism and its political legacy in Italy \citep[e.g.,][]{acemoglu2022war, fontana_historical_2018}. Relying on a unique natural experiment in the history of the world's largest democracy, I present evidence that authoritarianism in a democracy can have high political costs and persistent effects such as on voting behavior, political representation, and confidence in institutions.

This paper is also related to the growing cross-disciplinary literature on single-party dominance in democracy. Even in democratic countries where elections are free and fair, the same party sometimes continues to win for decades. Why does such single-party dominance endure longer in some places, and what leads to such parties' decline? Popular explanations for the decline of dominant parties include economic growth \citep{dasgupta_technological_2018}, economic crisis \citep{magaloni2006voting}, market liberalization \citep{greene2007dominant}, institutional change \citep{cox1997making}, party strategies \citep{levite1983legitimation}, weak institutions \citep{kohli1990democracy}, opposition coordination \citep{ziegfeld2017opposition}, opposition party organization combined with social cleavages \citep{tudor2019social}, and rise of minority groups in politics \citep{jaffrelot2003india} among others. I contribute to this growing literature by providing causal evidence and plausible mechanisms that a short-term policy implemented in the past undermining democratic rule can have lasting impacts on the decline of single-party dominance in politics. It has present-day implication as the short-term policy responses to the COVID-19 pandemic has exacerbated the situation undermining democratic rule in many countries, including single-party dominated democracies.

Finally, this paper also contributes to the literature on the Indian political economy and the evolution of political parties in India. For a long period, beginning from its independence in 1947, a single major party governed India. However, since the late 1970s, it has transitioned from single-party dominance to multiparty competition, particularly, coalition governments, with the formation of several regional parties. I contribute to the literature on change in the party structure \citep{chhibber1998party, chhibber_ideology_2018, dasgupta_technological_2018, ziegfeld_coalition_2012}, political cleavages \citep{banerjee_growing_2019}, and the evolution and popularity of the populist party in India \citep{kenny2017populism, suryanarayan_when_2019}. I present evidence that the authoritarian rule between 1975 and 1977 as a shock to Indian democracy has had a profound impact on Indian political economy. Additionally, the legacy of this brief period of authoritarian rule persists today.

The remainder of the paper is organized as follows. The next section provides the historical context. Section 3 describes the data used in the analysis and its sources. Section 4 discusses my empirical strategy to identify the causal effect. Section 5 presents some stylized facts and the main results. Section 6 explores the mechanisms. Section 7 examines the present-day consequences on confidence in politicians, and Section 8 concludes. The online Appendix provides additional robustness checks and results.

\section{The Emergency as a Shock to India's Democracy}

On June 25, 1975, Prime Minister Indira Gandhi of the INC party proclaimed a national emergency under Article 352 of the Indian constitution.\footnote{Article 352 (1) states that ``If the President is satisfied that a grave emergency exists whereby the security of India or of any part of the territory thereof is threatened, whether by war or external aggression or armed rebellion, he may, by Proclamation, make a declaration to that effect in respect of the whole of India or of such part of the territory thereof as may be specified in the Proclamation Explanation. A Proclamation of Emergency declaring that the security of India or any part of the territory thereof is threatened by war or by external aggression or by armed rebellion may be made before the actual occurrence of war or of any such aggression or rebellion, if the President is satisfied that there is imminent danger thereof.''}
The exact reason for the proclamation of emergency remains controversial to this day. However, political scientists, historians, and sociologists argue that a combination of political and economic problems facing the Prime Minister and India could be the most predictable factor.\footnote{For a detail overview of this period, please refer to \citet{jaffrelot2021india, dhar_indira_2018}, and \citet{nayar_emergency_2013}.}

The executive power associated with the proclamation of emergency allowed Mrs. Gandhi to suspend a wide range of civil liberties under the Indian constitution. Acting in an authoritarian manner, she misused this power, repressed the opposition, and instituted censorship in the name of law and order during this period. Hundreds of thousands of people, including leading opposition leaders, were arrested, the press censored, and public gatherings and strikes were declared illegal. With all the power in her hands as Prime Minister, Mrs. Gandhi undertook a series of new legislation and constitutional amendments to govern the country and extend the emergency period. Furthermore, she delayed the parliament and state elections several times \citep{gwatkin_political_1979}. In January 1977, however, Mrs. Gandhi unexpectedly called an election, released opposition leaders from jail, lifted press censorship, and once again permitted public meetings. The emergency period officially ended in March 1977 after the defeat of the INC party in the Lok Sabha election.

A hallmark policy synonymous with the emergency that directly affected the general population during this period was the aggressive family planning program involving forced sterilization.\footnote{Although India's family planning program existed prior to the emergency, the sterilization drive was intensified during this period.}
The program commenced in April 1976 with a new population policy introduced to the parliament by the Union Ministry of Health and Family Planning. The principal aim of the policy was to lower the population growth rate by boosting the family planning program. The new policy incorporated a series of fundamental changes to reduce population growth, including a substantial increase in monetary compensation for acceptors of sterilization, state-level introduction of incentives and disincentives attached to family planning policies, the disenfranchisement of states that failed to control fertility rates, allocation of central assistance to states according to family planning performance, and, most controversially, provisions for state governments to pass compulsory sterilization legislation \citep{singh_national_1976}.

With the introduction of the new population policy, the central government endorsed various coercion measures for sterilization and, in extreme cases, compulsory sterilization. The central and state governments increased the financial rewards for those who accepted sterilization. Through a range of incentives and disincentives, they pressured government employees to become sterilized and motivated others to do so. In several instances, quotas were imposed on central and state government employees to produce a certain number of individuals for sterilization. In other cases, citizens were required to produce sterilization certificates to access basic facilities, including housing, irrigation, subsidized foods through ration cards, and public health care. Other extreme measures were undertaken; for example, the state government in Maharashtra passed a bill allowing compulsory sterilization of couples with three or more children \citep{shah_commission_of_inquiry_third_1978}.\footnote{However, the bill was not approved by the central government and was eventually returned to the state for revision.}

The aggressive family planning program led to more than 8 million sterilizations between April 1976 and March 1977, more than three times the number in the previous year (see Figure \ref{figure:FigureA2}). Over 1.7 million sterilizations were performed during September 1976 alone, a figure that was equivalent to the annual average number of sterilizations performed during the preceding 10 years \citep{gwatkin_political_1979}. Most of the sterilizations performed during the period involved men undergoing vasectomies. Of the approximately 8.3 million sterilizations performed in 1976--1977, about 6.2 million (75\%) were achieved through vasectomies.

Survey evidence suggests that individuals were influenced or, in most cases, coerced into accepting sterilization during the forced sterilization period. In a survey of individuals who accepted sterilization during this period, \citet{panandiker_family_1978} document that about 72\% underwent sterilization due to the influence of government officials. Only about 19\% were sterilized on their own initiative, while the remaining 9\% were motivated by friends and relatives. Strikingly, no individuals were sterilized because of the lure of money (as a result of the increased compensation) and no one cited any case where money had played a motivating role. This evidence suggests that most individuals underwent sterilization involuntarily during this period.

The draconian forced sterilization during the emergency period was the first major program since independence under which Indian citizens directly suffered at the hand of the government. \citet{tarlo_body_2000} noted that the word ``emergency'' itself became synonymous with ``sterilization'' and, even today, individuals refer to the emergency period as the period of sterilization. The impact was such that the Indian government had to change the name of its ministry of health from the Department of Health and Family Planning to the Department of Health and Family Welfare in 1977 \citep{scott_my_2017}. In the post-emergency period, the family planning program shifted its focus from vasectomies to tubectomies, under which women emerged as the primary sterilization target \citep{basu_family_1985}. The emergency period remains controversial today and is regarded as one of the darkest periods in the history of Indian democracy.

Mrs. Gandhi misused the Indian constitution and proclaimed a national emergency. As a result, Indian citizens directly suffered at the hands of the government during this period. Therefore, this brief period of authoritarian rule may have generated a backlash against the central government in general and the political party (the INC) pursuing authoritarian rule in particular. Additionally, as an alternative definition of the authoritarian rule, my analysis tests for hypotheses examining whether states heavily affected by the draconian forced sterilization policy during the emergency period have different effects in terms of political representation and voting behavior in the subsequent elections.

\section{Data Sources and Description}

\subsection{Electoral Data}
The election data that I use are taken from all general election results in the Lok Sabha from 1962--2019 and the universe of state-level election results between 1962 and 2023. The data come from the statistical reports published by the Election Commission of India.\footnote{See https://eci.gov.in/statistical-report/statistical-reports/}
Recently, they have been digitized and harmonized by the Trivedi Center for Political Data, Ashoka University \citep{tcpd_ashoka}.\footnote{See http://lokdhaba.ashoka.edu.in/}
The data on election results since 1962 are available at both candidate and constituency levels.

To study the change in voting behavior toward the INC party, I focus on the election data at the candidate level. The data set contains the number of votes and the share of each candidate's vote in an election in a constituency. I rely on the percentage of the vote obtained by each candidate in a constituency as my primary outcome variable.\footnote{This is because the number of votes may provide biased results because the voting population varies substantially between constituencies.}
The data set also contains constituency-level data on the number of candidates contesting an election and the percentage of eligible voters who turn out to vote in each election. I use this information to study the mechanisms by examining contestants' behavior and voter turnout rates.

\subsection{Data on Family Planning During the Emergency}
The estimates on India's aggressive family planning policy during the emergency period come from the yearbooks published by the Ministry of Health and Family Planning, Government of India.\footnote{In 1977, the name of the department changed to Ministry of Health and Family Welfare,  Government of India.}
Along with various demographic and health statistics, the yearbook reports yearly statistics on family planning programs performed at the national and state levels from April in each year to March of the following year. Concerning the sterilization statistics, the yearbook reports the number of sterilizations performed each year at the state level. Additionally, it reports the number of vasectomy (male sterilization) and tubectomy (female sterilization) operations performed by each state each year.

I estimate the aggressive family planning policy by constructing two sterilization measures.\footnote{I primarily compare data from one year with data from the previous year to account for the emergency period because the emergency period continued from 1975 until 1977.}
First, as my baseline measure, I use all excess sterilizations (both male and female) performed during 1976--1977 compared with 1975--1976. Additionally, as an alternative measure, I consider the excess vasectomies performed during 1976--1977 compared with the 1975--1976 period because the forced sterilizations largely targeted male sterilization, with about 75\% of the sterilizations performed during this period involving vasectomies.

\subsection{Data on Confidence in Politicians}
My additional outcome variable to explore the present-day consequence of authoritarian rule is based on the data on confidence in institutions from the Indian Human Development Survey-II in 2011--2012 (IHDS-II) \citep{desai_national_2012}. The IHDS-II is a nationally representative survey that asks households questions about their confidence in politicians to fulfill promises. The respondents can choose between three answers: a great deal of confidence, some confidence, and hardly any confidence. The IHDS-II assigns values of 1, 2, and 3, respectively, to each of the answers.\footnote{Therefore, a higher score constitutes a lower level of confidence.} I use these data to explore whether the authoritarian rule in general and the forced sterilization during this period, in particular, have any durable consequences for confidence in politicians in India.

\section{Empirical Strategy}

I use five empirical strategies to present my main findings. My first strategy consists of providing descriptive evidence by comparing election outcomes between INC and non-INC candidates in a constituency across different periods before and after the emergency rule. To provide a visual representation, I present the raw data in a scatter plot. I then present the difference in means using simple two-by-two tables.

Second, I implement a DID strategy and present the results in a regression framework controlling for differences across regions, parties, and election years. My baseline estimation equation is:

\begin{equation}
Y_{icspt}=\beta Post_t \ast INC_p + \gamma_t + \delta \gamma_p + \gamma_c + \gamma_s + \epsilon_{icspt}
\end{equation}

where $Y_{icspt}$ denotes one of the election outcome measures of candidate i, contested from constituency c of state s, under party p, in year t. $Post_t$ takes a value of one for elections held after the emergency period (that is, 1977 and after), and a value of zero otherwise. $INC_p$ is an indicator variable for candidates belonging to the INC party. $\gamma_t$ are election year fixed effects. $\gamma_p$ represents party-level controls (in particular, party and party-type fixed effects). Because constituencies and states have changed over time, I use both constituency fixed effects $\gamma_c$ and state fixed effects $\gamma_s$. Standard errors are clustered at the constituency level from which a candidate contests electoral districts.

The main identifying assumption for my DID estimation approach above is that in the absence of the emergency rule, the evolution of the voting outcomes between the INC and other parties would have been similar. This may be a strong assumption, which may not be satisfied on several grounds. In particular, comparing outcomes between INC and non-INC candidates could be problematic because INC has been the largest political party in India since its independence in 1947.

To account for this issue, my third strategy consists of imposing a second level of variation in the authoritarian rule and implementing a DDD estimation strategy. I exploit the aggressive policy undertaken during this period, that is, the forced sterilization policy, and use the excess sterilizations performed at the state level as a second level of variation. More specifically, I estimate the following DDD specification:

\begin{equation}
Y_{icspt}=\beta Post_t\ast INC_p \ast S_s + \gamma_t + \delta \gamma_p + \gamma_c + \gamma_s + \mu' Interactions + \epsilon_{icspt}
\end{equation}

where $S_s$ is the number of excess sterilizations performed between April 1976 and March 1977 at state s normalized by performance in 1975--76. Specifically, I define $S_s$ as

$${Excess Sterilization}_s = \frac{\# sterilizations {(1976~1977)}_s - \# sterilizations {(1975~1976)}_s} {\# sterilizations {(1975~1976)}_s}$$
Additionally, $Interactions$ represents the interaction terms required to perform the DDD analysis. The other variables are the same as those defined in equation (1).

Fourth, I estimate a flexible DDD model that allows for the estimation of election year-specific treatment effects. To do this, I estimate the following regression:

\begin{equation}
Y_{icspt}=\sum_{t}\beta_tD_t \ast {INC}_p \ast S_s + \gamma_t + \delta \gamma_p + \gamma_c + \gamma_s + \mu' Interactions + \epsilon_{icspt},
\end{equation}

where $D_t$ is a dummy variable indicating whether the election belongs to year t.\footnote{The omitted category corresponds to the immediate major election year before the emergency (i.e., 1971).}
The other variables are the same as those defined in equation (2). The coefficient of interest is $\beta_t$, which captures the year-specific treatment effects. The pattern of $\beta_t$ in the pre-emergency period allows me to check the validity of the identification assumption. In the absence of any existing trends that correlate with excess sterilization, I expect to find no effects for pre-emergency periods, i.e., $\beta_t\approx0$.

Fifth and finally, I estimate a regression discontinuity design (RDD) estimation approach. Note that we can not conduct an RDD analysis of the Lok Sabha election data because of the problems associated with the frequency of running variable, which is the election year. To account for this issue, I use the universe of state-level election results between 1962 and 2023. This data set is particularly useful for RDD analysis as state-level elections frequently happen multiple times every year.\footnote{I use both year and month of election as running variables to conduct RDD analysis.}

\section{The Effects of Authoritarianism on Voting Behavior}

My main outcome variable for voting behavior is the share of votes held by candidates and by the winner in a constituency. As highlighted above, the 1977 election was held in March, after the emergency rule was relaxed in January. Hence, throughout, I interpret voting behavior in 1977 and after as belonging to the post-authoritarian period. Additionally, I consider the INC party candidates as my treatment group because the emergency was declared solely by Prime Minister Indira Gandhi of the INC party, and all the opposition leaders (including the senior leaders) were imprisoned during this period.\footnote{Anecdotal evidence suggests that Mrs. Gandhi declared the emergency without even consulting most of her INC party members. For an overview of this period and a detailed insider's account, see \citet{dhar_indira_2018} by the head of the Prime Minister's secretariat, who was one of her closest advisers during this period.}

\subsection{Descriptive Evidence}
I begin by presenting the raw data of the national election results through simple scatter plots in Figure \ref{figure:Figure3}.\footnote{In Figure \ref{figure:FigureA3} in the Appendix, I restrict the raw data to major election years only and present it for clear visualization.}
Panel A of Figure \ref{figure:Figure3} plots the average vote share of INC and non-INC candidates received in a constituency in each election year. In panel B, I additionally plot the probability of winning for INC and non-INC candidates in a constituency in each election year. The dashed line represents the end of the authoritarian rule in India. 

As we can observe, the average vote share and the probability of winning of INC candidates show an upward trend (or remains more or less flat) in the pre-emergency era but decline sharply in the post-emergency period. In contrast, the non-INC candidates' average share of votes and winning probability demonstrate a downward or constant trend in the pre-emergency era and remain flat in the post-emergency period.\footnote{The average share of votes and winning probabilities of INC and non-INC candidates do not sum to 100 and 1, respectively, because more than one non-INC candidate contested their electoral constituency.} Another interesting facts that we can observe from the raw data is that there was a discontinuous change in INC candidates' vote share and winning probability \emph{just} after the post-authoritarian period. However, we do not see such a sharp discontinuity in the case of non-INC candidates. In particular, we see a sudden decline in INC candidates' votes share and the probability of win just after the emergency rule in India. Overall, the figure reveals that the INC's share of votes and winning probabilities declined after the authoritarian rule, whereas those of the non-INC remain more or less flat.

\begin{center}
\centering Figure \ref{figure:Figure3} about here
\end{center}

Table \ref{figure:Table1} presents the levels and changes in vote share and winning probabilities using simple two-by-two tables. I present the results on vote shares in columns (1) and (2), and probabilities of winning in columns (4) and (5). In addition, I show the differences in average vote shares between INC and non-INC candidates in column (3) and the average winning probabilities between INC and non-INC candidates in column (6). Row (3) of the table presents the changes in average vote shares and winning probabilities before and after the emergency rule.

\begin{center}
\centering Table \ref{figure:Table1} about here
\end{center}

As Table \ref{figure:Table1} indicates, for INC candidates, both average vote shares and the winning probabilities declined after the emergency period, which is similar to the results documented in Figure \ref{figure:Figure3}. More interestingly, the non-INC share of votes and winning probabilities also declined.\footnote{We will explore the reason for these interesting phenomena in more detail in the mechanism section.} However, it declined more for INC candidates. The difference in these differences (see the estimates in column (3) row (3) and column (6) row (3)) suggest that, on average, INC candidates received 1.25 percentage points less votes, and their probability of winning was 27 percentage points lower. These estimates are highly statistically significant. This simple estimator can be interpreted as the effect of the authoritarian rule, under the assumption that in the absence of the emergency, the decline in voting share and the probability of winning would not have been systematically different between INC and non-INC candidates. In the remainder of this paper, I will elaborate on this assumption step by step in detail to provide a more convincing result.

\subsection{DID Estimates}
I present the DID estimates of equation (1) for vote shares in Table \ref{figure:Table2}. The coefficient in column (1) is my most parsimonious specification for performing the DID estimation in a regression framework. It includes only candidates' political party and year fixed effects. These fixed effects are included in all of my specifications, which ensures that the results are not driven by some unobserved political party characteristics and periodic trends over time. The post-INC variable (the DID term, henceforth) has a coefficient of -4.409 with a standard error of 0.568 (and is thus significant at less than 1\%). The coefficient estimate implies that the authoritarian rule is associated with declines in INC candidates' share of the votes by about 4.4 percentage points in the subsequent years.

\begin{center}
\centering Table \ref{figure:Table2} about here
\end{center}

The rest of the table shows that this relationship is robust when I control for a range of other covariates. In column 2, I include additional party-level controls (particularly types of political party fixed effects, which proxy for the geographical representation of a party for which candidates are contesting, such as national, state-based, local, and independent parties). The inclusion of these additional party-level fixed effects has virtually no effect on my coefficient estimates. In columns 3 and 4, I separately add state and constituency fixed effects, which control for permanent differences in political attitudes in the states and constituencies. In column 5, I include both state and constituency fixed effects. The inclusion of these geographic level controls, separately or together, has no discernible impact on the coefficient estimates for the DID term.

Finally, in column 6, I restrict my analysis to major elections only and exclude special elections (such as by-elections) held between general elections to fill vacant positions. The coefficient estimates for the DID term in column 6 do not change and are similar to those presented in column 5. My overall interpretation of the results in Table \ref{figure:Table2} is that the authoritarian rule had a first-order impact on the decline of the INC's share of votes in the subsequent elections.

Next, I present the DID estimates of the probability of winning the elections in Table \ref{figure:Table3}. In addition to validating my results using shares of votes, this measure is of interest because it directly captures the political outcome in constituencies that oppose the INC and because it is informative about the evolution of political parties in India and how the authoritarian rule has impacted the decline of the INC's dominance.

\begin{center}
\centering Table \ref{figure:Table3} about here
\end{center}

Table \ref{figure:Table3} has a similar structure as Table \ref{figure:Table2}. In all six columns of Table \ref{figure:Table3}, we see a sizable impact of the authoritarian rule on the subsequent decline in the winning probability of INC candidates. In my most parsimonious specification in column 1 (which includes party fixed effects and election year fixed effects as in column 1 of Table \ref{figure:Table2}), the coefficient estimate is -0.276, with a standard error of 0.017. This magnitude implies that the authoritarian rule is associated with declines in INC candidates' probability of winning the elections in the subsequent years by about 28 percentage points. Hence, the authoritarian rule between 1975 and 1977 accounts for a significant portion of the decline in the INC's political dominance in India.

The estimates in the remaining columns are stable. Columns 2 and 3 add party-type fixed effects and state fixed effects, but the coefficient does not change. Column 4 includes constituency fixed effects in place of state fixed effects, but the coefficient varies only slightly (from -0.276 to -0.274). Columns 5 and 6 present my most demanding specification, including all the controls in column 5 and excluding special elections in column 6, suggesting that my results are robust and fairly stable to alternative specifications.

\subsection{DDD Estimates}
Now, I turn to the DDD estimation approach using a second level of variation in the authoritarian rule. I exploit a distinctive, aggressive policy undertaken during this period and use the number of excess sterilizations performed between April 1976 and March 1977, normalized by the previous year's performance, to implement my DDD estimation. As noted previously, among all authoritarian policies, aggressive family planning through forced sterilization was a hallmark of the emergency period that directly affected the general population. This measure is of particular interest because it uniquely captures the INC's authoritarian policy at the height of the emergency rule and because it is informative in capturing some of the legacies that might have directly impacted later support for the INC party.

Table \ref{figure:Table4} presents the DDD results exploiting this source of variation. Panel A presents the DDD estimates of vote shares. In panel B, I present the DDD estimates of the probability of winning the elections. Table \ref{figure:Table4} has an identical structure to Table \ref{figure:Table2} and \ref{figure:Table3}. The results in this table are uniformly consistent with our hypothesis that the authoritarian rule, as captured by excess sterilization, has a large and statistically significant effect on the INC's election outcomes in subsequent years. For example, the estimate for the share of votes in panel A of Table \ref{figure:Table4} in our most demanding specification in column 6 is -0.874 (with a standard error of 0.313), which implies that constituencies with a mean level of excess sterilization (about 3.38 times) are associated with a decline of about 2.95 percentage points in the INC candidates' shares of votes in the subsequent years. The estimate for the probability of a win in the same specification is -0.035 (with a standard error of 0.009) and implies a sizable negative effect as well: a constituency with a mean level of excess sterilization is associated with a decline of 11.86 percentage points in the INC candidates' probability of winning the elections in the subsequent years.

\begin{center}
\centering Table \ref{figure:Table4} about here
\end{center}

\subsection{Flexible DDD Approach: Event Study Design}
I then turn to the flexible DDD approach described in equation (3) that allows for the estimation of election year-specific treatment effects. I present the $\beta_t$ coefficients through event study graphs in Figure \ref{figure:Figure4}. Additionally, I present the estimated coefficients in tabular form in Table \ref{figure:TableB1} in the Appendix. For better visualization, I present the coefficient estimates for major elections only and exclude special elections (such as by-elections or by-pools) held between general elections, as there are few observations in these years.\footnote{For reference, I present the coefficient estimates for all elections in Figure \ref{figure:FigureA4} in the Appendix.}

Panel A of Figure \ref{figure:Figure4} plots year-specific estimates for vote shares. The black dots show the main impact, with 95\% confidence intervals represented by blue lines. As we observe, the estimates of vote shares for pre-emergency periods, i.e., the pre-1977 years, in panel A of Figure \ref{figure:Figure4} fluctuate around zero and do not follow any specific trend. This finding constitutes a validity check on my main identification assumption of a lack of existing trends. However, the coefficients for the post-emergency years are negative, except for 1984, and most years are highly statistically significant. The positive coefficient of the 1984 election remains unclear and needs further investigation. However, the most likely reason is that the 1984 election was held immediately after the assassination of Prime Minister Indira Gandhi, who declared the emergency and instituted authoritarian rule in India.\footnote{Indira Gandhi was assassinated on October 31, 1984, and the general election was held in December 1984. Most of India voted for the INC during the 1984 election, and it is the only time in history in which a single party won more than 400 seats.}
The results in panel A of Figure \ref{figure:Figure4} suggest that authoritarian rule accounts for a persistent decline in the INC's vote share and that the effect remains up to the recent election in 2019.

\begin{center}
\centering Figure \ref{figure:Figure4} about here
\end{center}

Next, I present the year-specific estimates of the probability of winning the elections in panel B of Figure \ref{figure:Figure4}. Panel B has a similar structure to the vote shares shown in panel A and it is evident that the figures in both panels are quite similar. Almost all coefficients for the post-emergency years remain negative. However, a major difference is that the estimates of the probability of winning for the pre-emergency years sometimes remain positive and statistically significant (although we do not observe any specific trend).

My overall interpretation of Figure \ref{figure:Figure4} is that the authoritarian rule between 1975 and 1977 had a first-order impact on the decline in the INC's political dominance in India. Additionally, the legacy of the authoritarian rule persists and continues to affect the INC's political representation in the long run.

\subsection{Regression Discontinuity Design Approach}

As a final step in my empirical estimation strategy, I examine the impact of authoritarian rule in India through a regression discontinuity design (RDD) approach. As explained above, we can not conduct an RDD analysis of the national-level election data because it does not happen every year. For this, I focus on state-level election data (Assembly Elections) between 1962 and 2023 which happens frequently multiple times every year. 

I present the RDD results in Figure \ref{figure:Figure5}. Panel A of Figure 5 presents the change in INC candidates' vote share. As we can observe, there is a discontinuity in the vote share of INC candidates at the cutoff point. Precisely, using election years as a running variable with alternative bandwidths, I find that the INC’s vote share, on average, is about 5-22 percentage points lower after the authoritarian rule. Using election months instead of election years as a running variable in my RDD specification also produces identical results (see Figure \ref{figure:FigureA5} in the appendix)

\begin{center}
\centering Figure \ref{figure:Figure5} about here
\end{center}

Next, I present the results of INC candidates' winning probability in Panel B of Figure 5. Panel B has an identical structure to Panel A. Similar to the vote share, I also document a sharp discontinuity in the winning probability of INC candidates after the emergency rule. The RDD estimation in panel B suggests that the emergency rule can explain between 17-56 percentage points drop in the INC candidates’ probability of winning state-level elections in subsequent years. These results from state-level elections further bolster my interpretation that the authoritarian rule led to a sharp decline in the then-dominant incumbent, the Indian National Congress party's political dominance in subsequent years.

\subsection{Robustness}
Further robustness checks for the results in this section are provided in the Appendix. I briefly discuss them here. In Table \ref{figure:TableB2}, I show that the results of my DDD estimates are robust to considering an alternative measure of aggressive policy undertaken during this period (forced sterilization) measured by excess vasectomy, which constituted the majority of sterilization operations (about 75\%) during this period.

As a final robustness check, I restrict my analysis to \textsl{ INC candidates only} and conduct a flexible DID estimation approach considering excess sterilization interacted with election years. In particular, I estimate the following event-study specification:

\begin{equation}
Y^{INC}_{icst}=\sum_{t}\beta_tD_t \ast S_s + \gamma_t + \gamma_c + \gamma_s  + \epsilon_{icst}
\end{equation}

where $Y^{INC}_{icst}$ denotes one of the voting outcome measures of INC candidate i, contested from constituency c of state s, in year t. The other variables are similar to those defined in equation (3). The coefficient of interest is $\beta_t$, which captures the year-specific treatment effects. I present the results in  Figure \ref{figure:FigureA6} in the Appendix. As we can see, the results are quite similar to the evidence shown in Figure \ref{figure:Figure4}.
\section{Understanding the Mechanisms}

In the previous section, I showed that the authoritarian rule instituted between 1975 and 1977 directly impacted the decline in the INC's political dominance in India. In this section, I investigate the possible mechanisms. In particular, I first show that the voter turnout rate did not decline after the authoritarian rule, suggesting that lower voter turnout did not cause a decline in the incumbent's share of votes in subsequent elections. Additionally, I provide suggestive evidence that the authoritarian rule gave rise to the formation of new political parties.

\subsection{Where Did the INC Votes Go?}
As a first step in building up the evidence for my mechanism, I check whether there is any change in the overall voter turnout rate. A possible direct mechanism is that the INC party's share of votes declined because many of their supporters who voted for them before the authoritarian period did not turn up to vote in subsequent years. I address this issue directly by examining the percentage of eligible voters who vote in a constituency. I present the raw data in Figure \ref{figure:Figure6}. Figure \ref{figure:Figure6}, which plots the average voter turnout rate in a constituency, shows that the voter turnout rate did not decline. Instead, the voter turnout rate increased after the end of the authoritarian rule.

\begin{center}
\centering Figure \ref{figure:Figure6} about here
\end{center}

Next, I implement a DID strategy and present the results in a regression framework, controlling for differences across regions and election years and considering a variation in the authoritarian rule measured by excess sterilization. My baseline estimation equation is:

\begin{equation}
Y_{cst}=\beta Post_t\ast S_s+\gamma_t+\gamma_c+\gamma_s+\epsilon_{cst},	
\end{equation}

where $Y_{cst}$ denotes the voter turnout rate in constituency c of state s in year t. The other variables are the same as those defined in equation (2). I present the results in Table \ref{figure:Table5}. Columns 1--4 present the coefficient estimates with excess sterilization as the measure of authoritarian rule. Columns 5--8 present the estimates with excess vasectomy as an alternative measure of authoritarian rule.

\begin{center}
\centering Table \ref{figure:Table5} about here
\end{center}

The results from Table \ref{figure:Table5} support the evidence shown in Figure \ref{figure:Figure6} that the voter turnout rate did not decline. Indeed, the relationship is positive, although not statistically significant, in every specification. Thus, this suggests that earlier INC supporters did not refrain from casting their vote but rather switched away from the INC and toward non-INC candidates in the post-authoritarian period.

\subsection{Evolution of New Political Parties}
As an alternative interpretation of my proposed mechanism, I examine whether the authoritarian rule led to the formation of new political parties. This insight is motivated by the results presented in Table \ref{figure:Table1}. As the Table \ref{figure:Table1} shows, the share of votes and winning probability declined in the post-emergency period for both the INC and non-INC parties. These findings are counter-intuitive because, in the absence of any change in the number of candidates contesting in a constituency, the decline in INC voting outcomes should cause the non-INC candidates' share of votes and winning probability to increase on average. However, I find the opposite effect. I explore the intuition behind this result in detail in this subsection.

The INC was the largest political party when authoritarian rule was imposed in India in the 1970s. From the first election in 1951--1952 after India become a republic, the INC party was constantly in power. Moreover, there were few other political parties during this period and, in particular, no strong opposition parties.\footnote{For example, the INC received 352 seats in the 1971 Lok Sabha election, the major election immediately before the authoritarian rule. The main opposition parties, the Communist Party of India (Marxist) (CPI(M)) and the Communist Party of India (CPI), received only 25 and 23 seats, respectively.}

My alternative interpretation is that the authoritarian rule and the draconian policy imposed during the emergency may have created a credibility gap in the INC's ability to rule India, which led to the formation of new opposition parties in direct opposition to the INC rule, which then captured the INC votes.

I provide direct evidence of my interpretation. I consider the number of candidates contested in an election from a constituency as a proxy measure for the number of political parties in a constituency. I plot the average number of candidates contested from a constituency in every election year in panel A and major election years in panel B of Figure \ref{figure:Figure7}. It is evident from the raw data that there was a sudden increase in the number of candidates in the post-authoritarian period. In particular, there was a sudden rise in the number of candidates after the 1977 election. The 1977 election was announced unexpectedly in January during the authoritarian rule and was held less than two months later, in March. Hence, there was a short lag before the increase in non-INC candidates because there was very little time to form new political parties and stand candidates for the election in 1977.

\begin{center}
\centering Figure \ref{figure:Figure7} about here
\end{center}

Next, I implement a DID strategy (similar to equation 5) and present the results in Table \ref{figure:Table6}. This table has an identical structure to Table \ref{figure:Table5}. The results from Table \ref{figure:Table6} support the evidence presented in Figure \ref{figure:Figure7} and my interpretation suggesting that the draconian sterilization policy as a measure of the authoritarian rule had a positive impact on the number of candidates contesting for election in the post-authoritarian period.

\begin{center}
\centering Table \ref{figure:Table6} about here
\end{center}

I provide additional support for this interpretation by conducting an event study analysis. In Figure \ref{figure:Figure8}, I conducted a flexible DID approach and present the election year-specific treatment effects through event study graphs. Figure \ref{figure:Figure8} plots the year-specific treatment effects considering excess sterilization as the measure of authoritarian rule.\footnote{the results are identical if I conduct an event study analysis considering excess vasectomy as my measure of authoritarian rule.} Overall, the consistency in the pattern of the findings bolsters my interpretation that the decline in the INC's share of votes was plausibly caused by the formation of new opposition parties as a direct deterrence to the INC rule.\footnote{We see a sudden increase in the number of candidates in 1996 election which is a special case in the history of Indian politics. The 1996 election was preceded by major scandals erupted against the incumbent as well as the opposition party members (such as `Jain hawala', `Tandoor murder case'). About 14000 candidates contested for the 1996 election, and it is the only time in India's history where more than 10000 candidates have contested for election in a parliamentary election. Excluding the 1996 parliament election also provide similar results.}

\begin{center}
\centering Figure \ref{figure:Figure8} about here
\end{center}

\section{Present-Day Consequences}

In this section, I investigate whether there are any durable political consequences of authoritarian rule in India. In particular, I focus on present-day confidence in politicians. I use data from the IHDS-II in 2011--2012 on confidence in institutions \citep{desai_national_2012}. Figure \ref{figure:Figure9} presents the association between excess sterilizations as a measure of authoritarian rule and confidence in politicians through scatter plots. We see a positive association in the raw data, suggesting that households located in states heavily exposed to sterilization during the authoritarian period exhibit a lower level of confidence in politicians.\footnote{Recall that the IHDS-II assigns values of 1, 2, and 3 to the responses ``a great deal of confidence,'' ``only some confidence,'' ``hardly any confidence at all,'' respectively. Therefore, a higher score constitutes a lower level of confidence.}

\begin{center}
\centering Figure \ref{figure:Figure9} about here
\end{center}

Next, I examine this relationship in a regression framework controlling for household and geographic characteristics in Table \ref{figure:Table7}. The coefficient in column (1) is the most parsimonious specification, where I estimate the association without any controls. The coefficient estimate is positive and similar to the association I found in Figure \ref{figure:Figure9}. The rest of the table shows that this relationship is robust when I control for a range of other covariates. In column 2, I include household-level controls. The inclusion of these controls has hardly any effect on the coefficient estimate. In column 3, I additionally include geographic controls. Again, there is no discernible impact on the coefficient estimate. Overall, I find a consistent and sizable negative effect of excess sterilization in states during the authoritarian rule period on present-day confidence in politicians.

\begin{center}
\centering Table \ref{figure:Table7} about here
\end{center}

My overall interpretation of these results is as follows: the authoritarian rule and draconian sterilization policy implemented by the \textit{then-largest political party} in India reduced confidence in the credibility of politicians on a long-term basis. Thus, the legacy of this historical event persists today.\footnote{Note that the evidence presented here is cross-sectional and that the survey was conducted in 2011--2012 when the INC held power in the central government. This should be kept in mind when interpreting these findings.}

\section{Conclusion}

In this paper, I examined the political consequences of authoritarianism in the context of India, the world's largest democracy. The state of emergency rule implemented by the then Prime Minister Indira Gandhi of the INC party between 1975 and 1977 was the first brush with authoritarian rule in India since independence in 1947. The unexpected nature of the emergency rule provides a unique natural setting in history to study the legacy of authoritarianism in a democracy.

Using national and state-level election data sets, I documented that the authoritarian rule had a first-order impact on the decline of the then-incumbent party's political dominance in India, with a sharp decline in the share of votes for the INC candidates in subsequent elections. I provided support for this interpretation by using several alternative estimation strategies, including DID, DDD, an event study, and an RDD approach.

The decline in the incumbent party's share of votes and winning probability was not caused by a lower voter turnout rate. I found that the voter turnout rate did not decline. Instead, there was a sharp rise in the number of opposition candidates contesting for election in subsequent years. This suggests that the authoritarian rule gave rise to the formation of new political parties in the following years and played an important role in the decline of the INC. Finally, I examined the durable consequences and found that households in states heavily exposed to the draconian policy implemented during this period exhibit lower confidence in politicians today. Thus, the enduring legacy of this historical event persists in the long-run.

A key question is how generalizable this historical episode in India is to other contexts. The most direct parallel is to countries that have faced or are currently facing extreme levels of democratic backsliding. Authoritarian regimes in democracies are not uncommon. For example, many Latin American democratic countries have had at least some experience with authoritarian governments since World War II. Fujimori's government in Peru in the 1990s is a prominent example that plausibly aligns with the Indian situation. Other countries facing extreme democratic breakdown include Turkey and Hungary, where the current governments have consolidated authoritarian power through constitutional amendments, and the Philippines, where the recent past regime was engaged in a dictatorial policy of war on drugs. Poland's right-wing polarization on religious and gender grounds and Brazil's use of dictatorship-era law to detain and investigate critics during the Bolsonaro's regime are two other examples. Figure \ref{figure:FigureA7} in the Appendix presents the decline in the democracy index in these countries.

What are the lessons for India today? The recent democratic backsliding in the world's largest democracy is of grave concern. The Variety of Democracy Institute has classified India as an ``electoral autocracy'' since 2019, and Freedom House downgraded India from ``free'' to a ``partly free'' country in 2021 \citep{hellmeier2021state, Freedom_house2021}. The findings from this paper have important implications for India as the current regime is pursuing authoritarian policies, including the introduction of the Citizenship Amendment Act in 2019 and the misuse of sedition law to crack down on freedom of expression. Perhaps more importantly, the findings presented here have present-day policy implications as Uttar Pradesh, the most populous state in India, proposed a controversial population policy in July 2021, and the draft bill for population control ``primarily through sterilization,'' coercion, incentives, and disincentives are currently being discussed.\footnote{A similar population policy is being discussed in Assam, a state in the northeastern part of India.}

\clearpage
\bibliography{autocracy_democracy.bib}
\bibliographystyle{chicago}


\clearpage
\centering
\textbf{FIGURES AND TABLES}

\clearpage


\clearpage
\begin{figure}[htbp]
\begin{center}
\caption{\label{figure:Figure3}\textbf{The Effects of Authoritarianism on Voting Behavior}}
\subcaption{Panel A: The Effects of Authoritarianism on Vote Share}
\includegraphics[height=8cm]{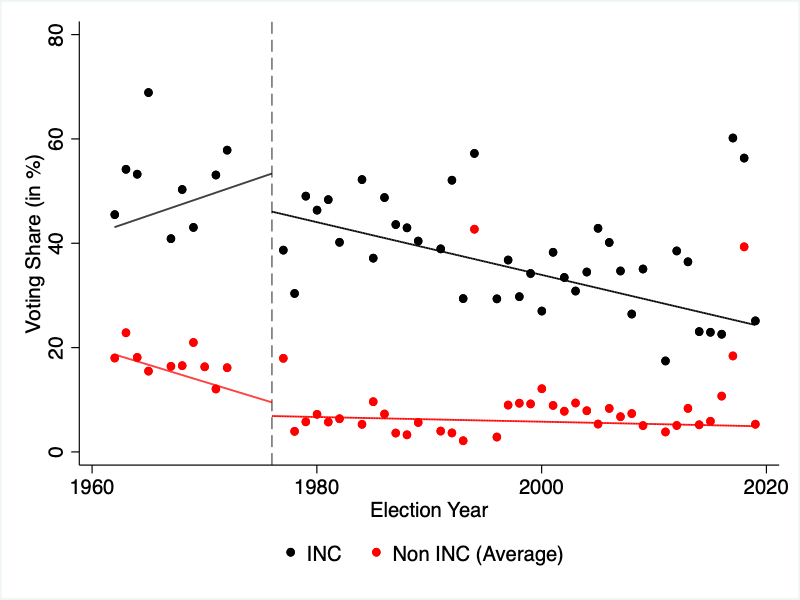}
\subcaption{Panel B: The Effects of Authoritarianism on Probability of Win}
\includegraphics[height=8cm]{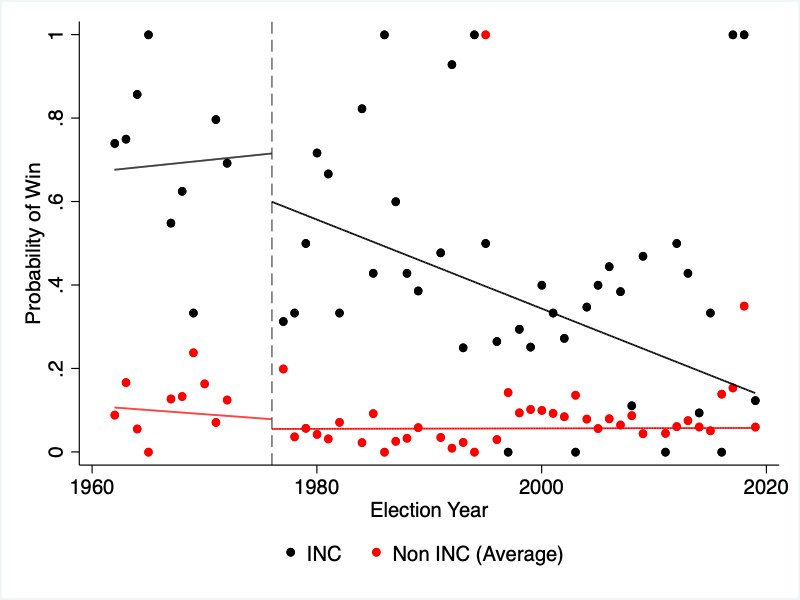}
\end{center}
{\small Notes: The figure plots the effect of authoritarianism on voting behavior and election outcomes in India. The data are from all the general elections to the lower house of the Indian parliament (Lok Sabha) between 1962 and 2019. Panel A plots the average vote share of INC and non-INC candidates received in a constituency in each election year. Panel B plots the probability in which INC and non-INC candidates won in a constituency in each election year. The dashed line represents the end of the authoritarian rule in India. The average of INC and non-INC voting share and probability of win does not sum to 100 and 1 respectively, because there are more than one non-INC candidates contesting for election in a constituency.
}
\end{figure}


\clearpage
\begin{figure}[htbp]
\begin{center}
\caption{\label{figure:Figure4}\textbf{DDD Estimates: Event Study (Major Election Years) }}
\subcaption{Panel A: Vote Share}
\includegraphics[height=8cm]{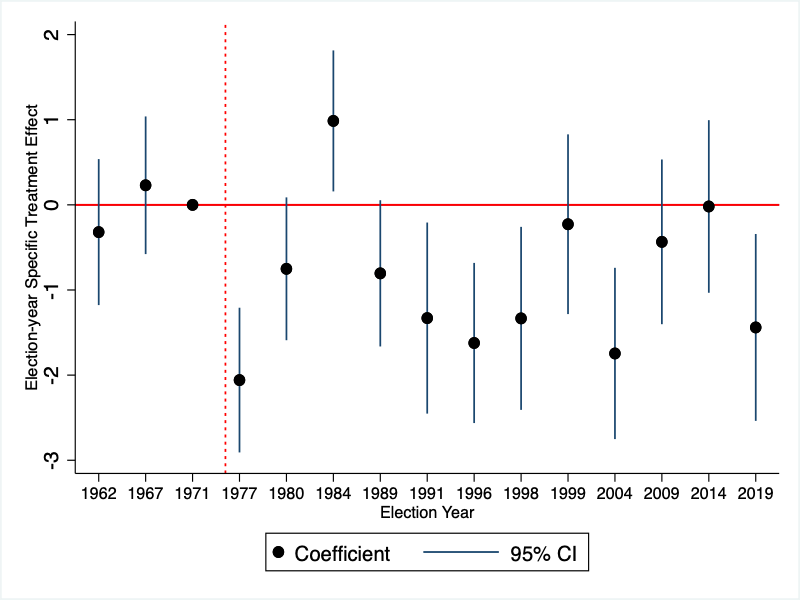}
\subcaption{Panel B: Probability of Win}
\includegraphics[height=8cm]{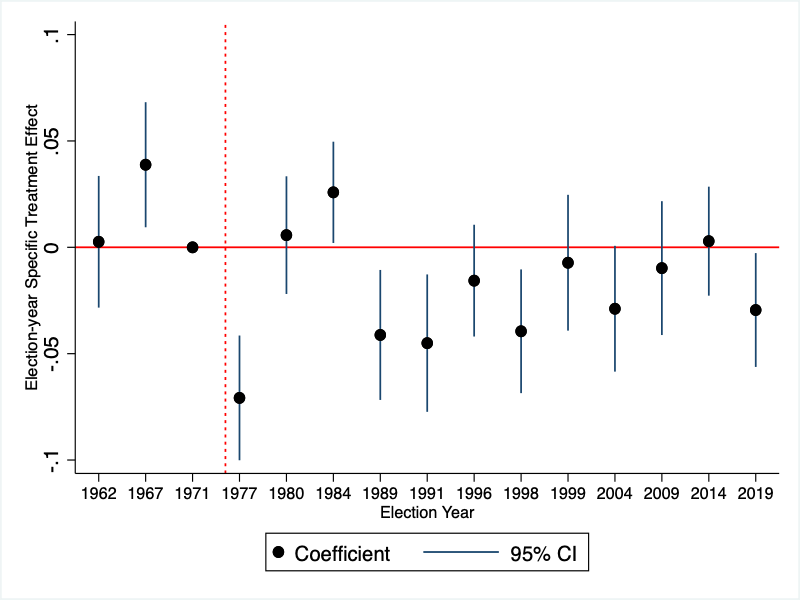}
\end{center}
{\small Notes: The figure presents the political consequences of authoritarianism in India through an event study framework. The data are from all major general elections to the lower house of the Indian parliament (Lok Sabha) between 1962 and 2019. Panel A plots the year-specific treatment effects of INC candidates' vote share. Panel B presents the year-specific treatment effects of INC candidates' probability of winning an election. The (red) dashed line represents the end of the authoritarian rule in India.
}
\end{figure}

\clearpage
\begin{figure}[htbp]
\begin{center}
\caption{\label{figure:Figure5}\textbf{Regression Discontinuity Design (RDD) Estimates}}
\subcaption{Panel A: INC's Vote Share in Assembly Election}
\includegraphics[height=8cm]{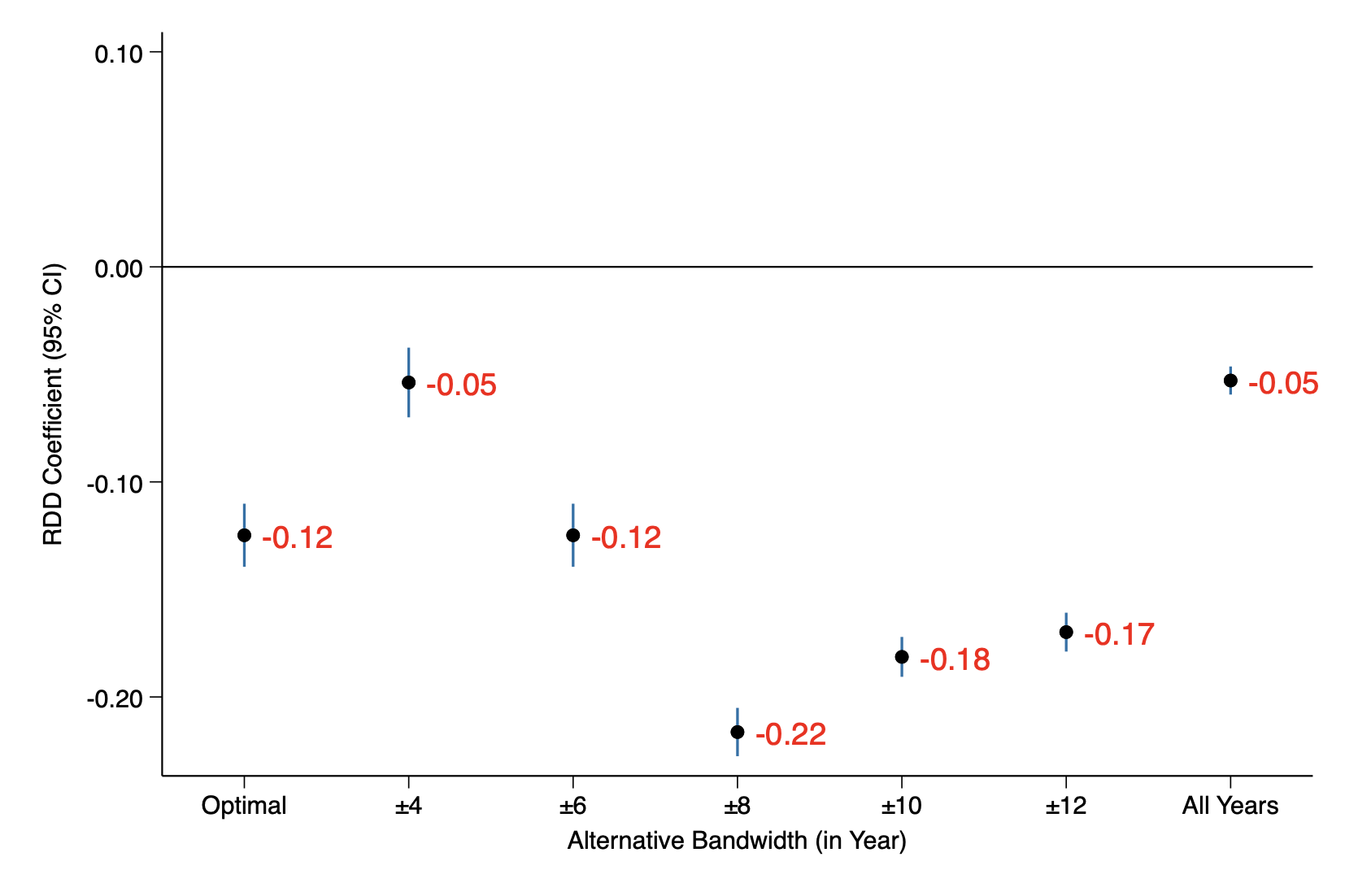}
\subcaption{Panel B: INC's Probability of Win in Assembly Election}
\includegraphics[height=8cm]{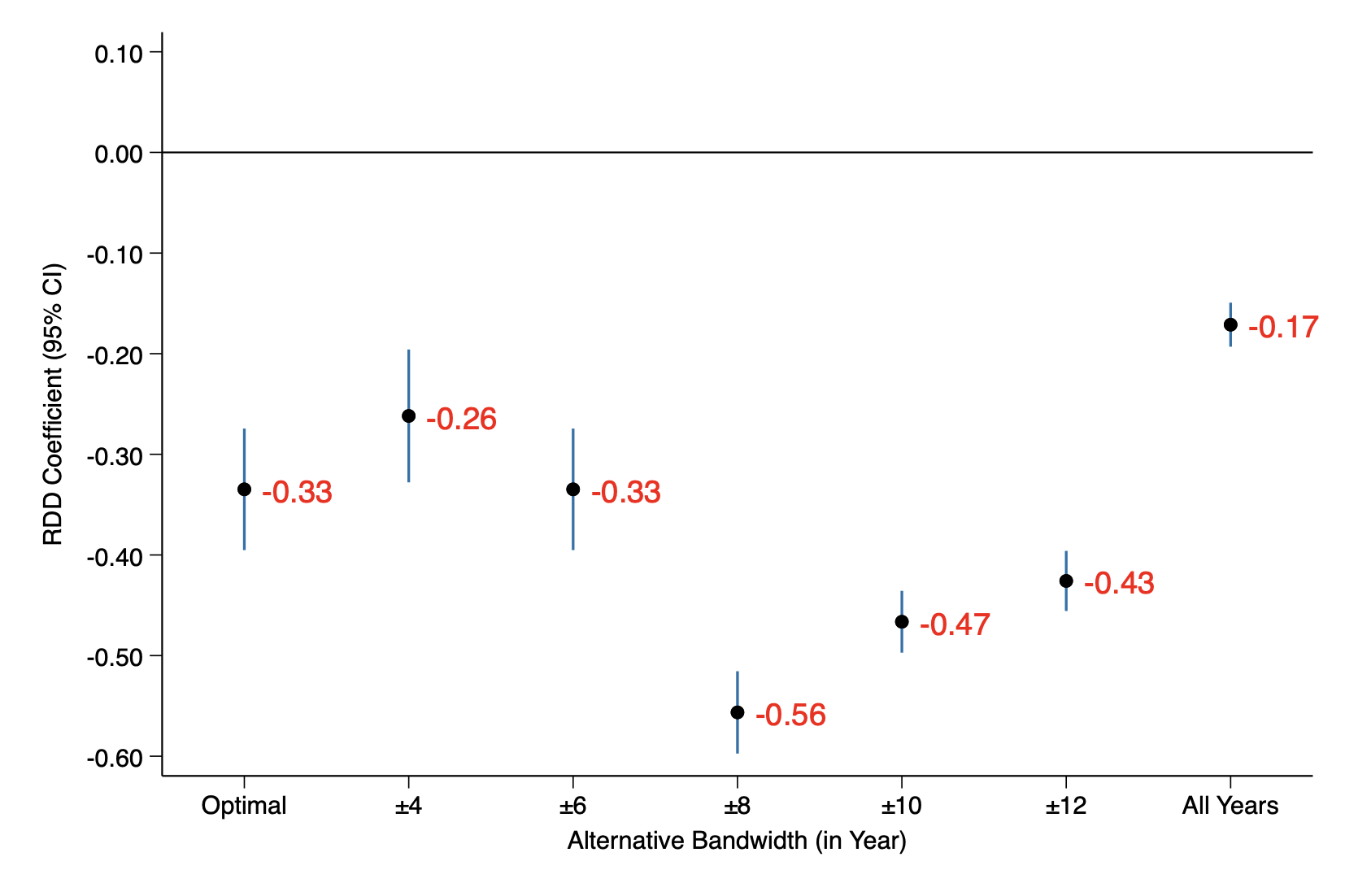}
\end{center}
{\small Notes: The figure presents RDD estimates. The data are from
the state-level Assembly Elections between 1962 and 2023. The running variable is the election year. Panel A plots the RDD coefficients of INC's Vote Share with alternative bandwidths. Panel B plots the RDD coefficients of INC candidates' probability of winning assembly election with alternative bandwidths. I set the local polynomial of order 1, which is simple, transparent, and easy to interpret (using local polynomial of order 2 also produces similar results). I use a uniform kernel (using a triangular kernel also produces identical results). Using election months (instead of election years) as a running variable produces identical results as well (See Figure \ref{figure:FigureA5} in the appendix).  }
\end{figure}


\clearpage
\begin{figure}[htbp]
\begin{center}
\caption{\label{figure:Figure6}\textbf{Average Voter Turnout Rate}}
\subcaption{Panel A: All Election Years}
\includegraphics[height=8cm]{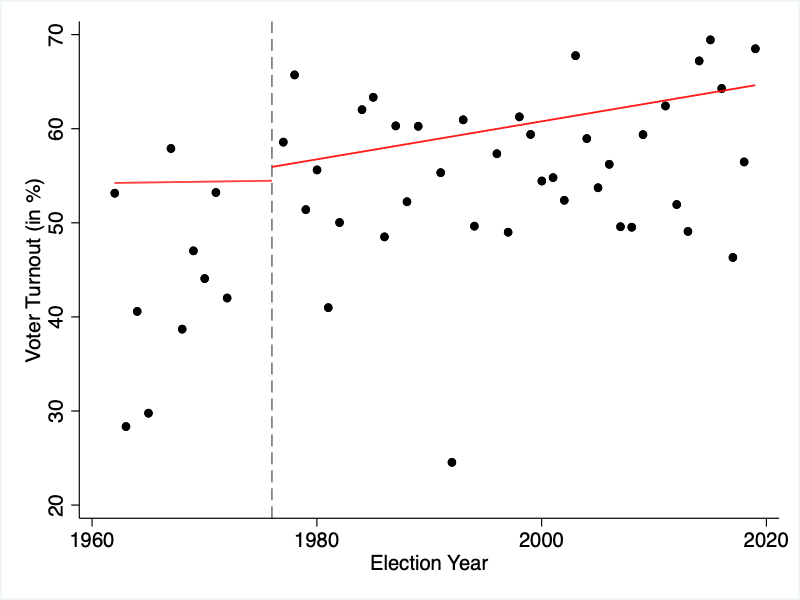}
\subcaption{Panel B: Major Election Years}
\includegraphics[height=8cm]{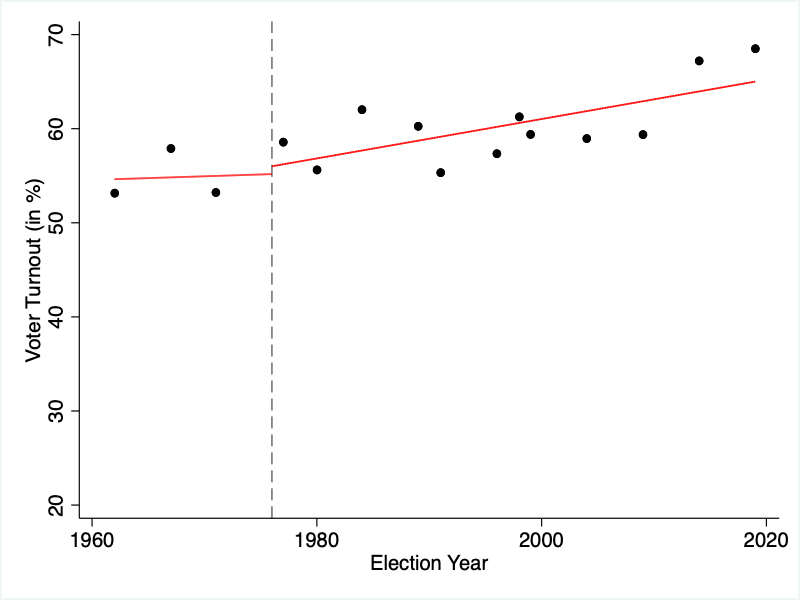}
\end{center}
{\small Notes: The figure presents the evolution of the voter turnout rate in India. The data are from the general elections to the lower house of the Indian parliament (Lok Sabha) between 1962 and 2019. Panel A plots the percentage of eligible voters who turn out to vote in a constituency in each election year. Panel B plots the percentage of eligible voters who turn out to vote in a constituency in major election years. The dashed line represents the end of the authoritarian rule in India.
}
\end{figure}

\clearpage
\begin{figure}[htbp]
\begin{center}
\caption{\label{figure:Figure7}\textbf{Average Number of Candidates Contested}}
\subcaption{Panel A: All Election Years}
\includegraphics[height=8cm]{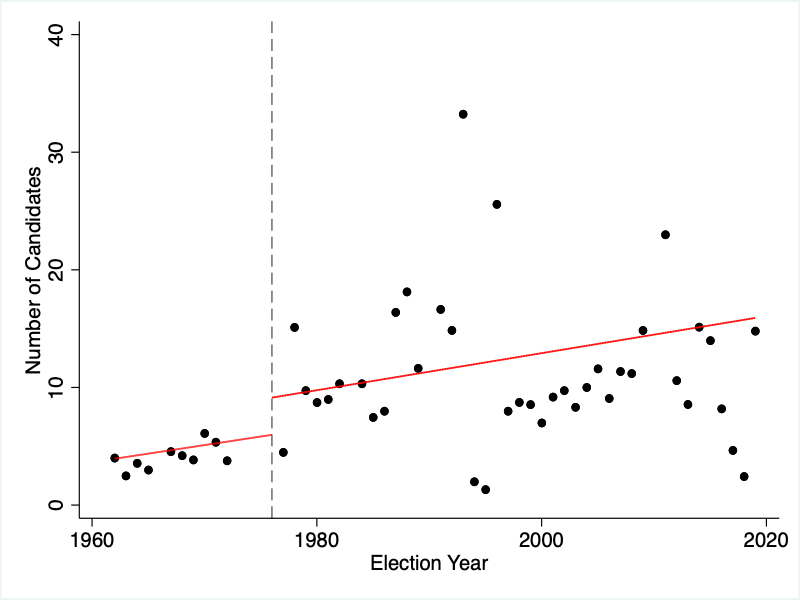}
\subcaption{Panel B: Major Election Years}
\includegraphics[height=8cm]{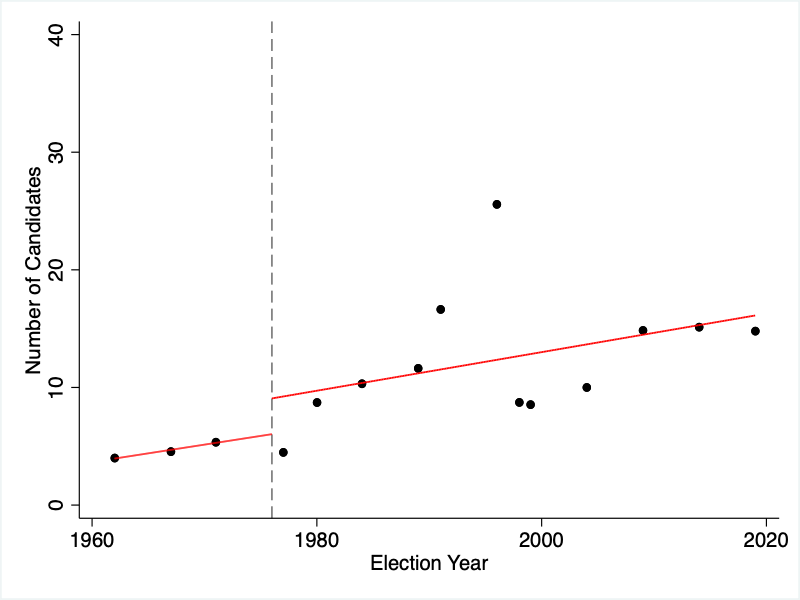}
\end{center}
{\small Notes: The figure presents the evolution of new political parties in India. The data are from the general elections to the lower house of the Indian parliament (Lok Sabha) between 1962 and 2019. Panel A plots the average number of candidates contested from a constituency in each election year. Panel B plots the average number of candidates contested from a constituency in major election years. The dashed line represents the end of the authoritarian rule in India.
}
\end{figure}


\clearpage
\begin{figure}[htbp]
\begin{center}
\caption{\label{figure:Figure8}\textbf{Event Study - Number of Candidates Contested}}
\includegraphics[width=\textwidth]{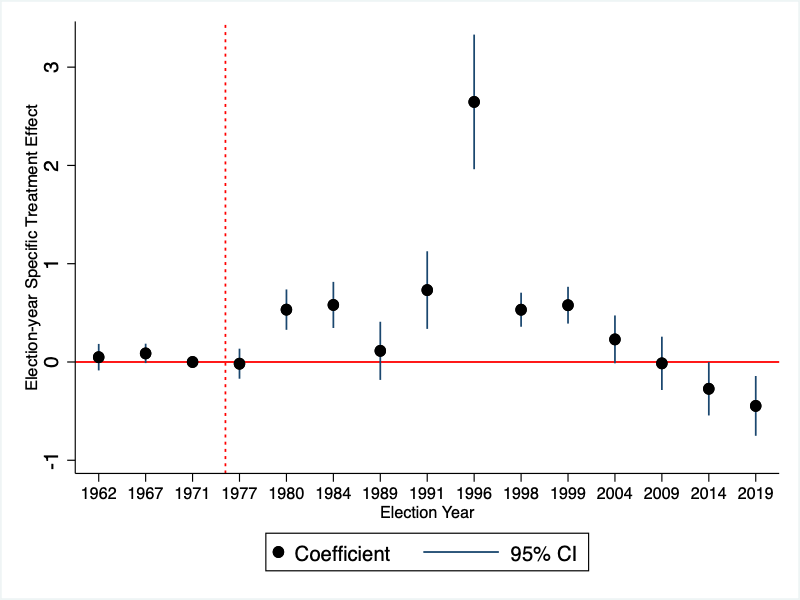}
\end{center}
{\small Notes: The figure presents the impact of authoritarianism on the evolution of new political parties in India. The data are from all major general elections to the lower house of the Indian parliament (Lok Sabha) between 1962 and 2019. The figure plots the year-specific treatment effects considering excess sterilization as a measure of authoritarian rule. The (red) dashed line represents the end of the authoritarian rule in India.
}
\end{figure}


\clearpage
\begin{figure}[htbp]
\begin{center}
\caption{\label{figure:Figure9}\textbf{Association Between Excess Sterilizations and Confidence in Politicians}}

\includegraphics[width=\textwidth]{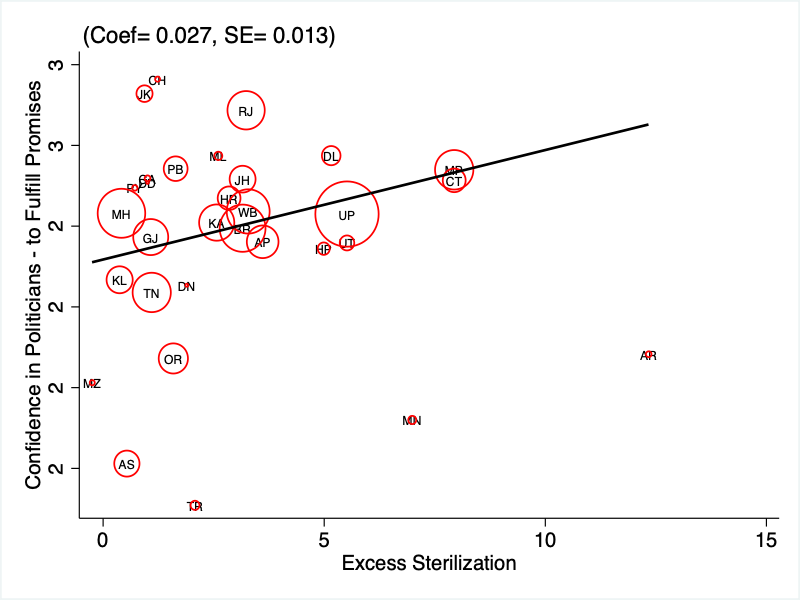}
\end{center}
{\small Notes: The figure presents the correlation plots of the present-day consequences. It plots the association between excess sterilizations in 1976--1977 and confidence in politicians. The data on confidence in politicians are from IHDS-II. It assigns the value 1 to ``a great deal of confidence,'' 2 to ``only some confidence,'' and 3 to ``hardly any confidence at all.'' (Therefore, a higher score constitutes a lower level of confidence.) The fitted lines are weighted by the population of the state and union territory.
}
\end{figure}


\clearpage
\begin{table}[htbp]
\begin{center}
\caption{\label{figure:Table1}\textbf{Average Vote Share and Probability of Win Before and After the Emergency}}
\includegraphics[width=\textwidth]{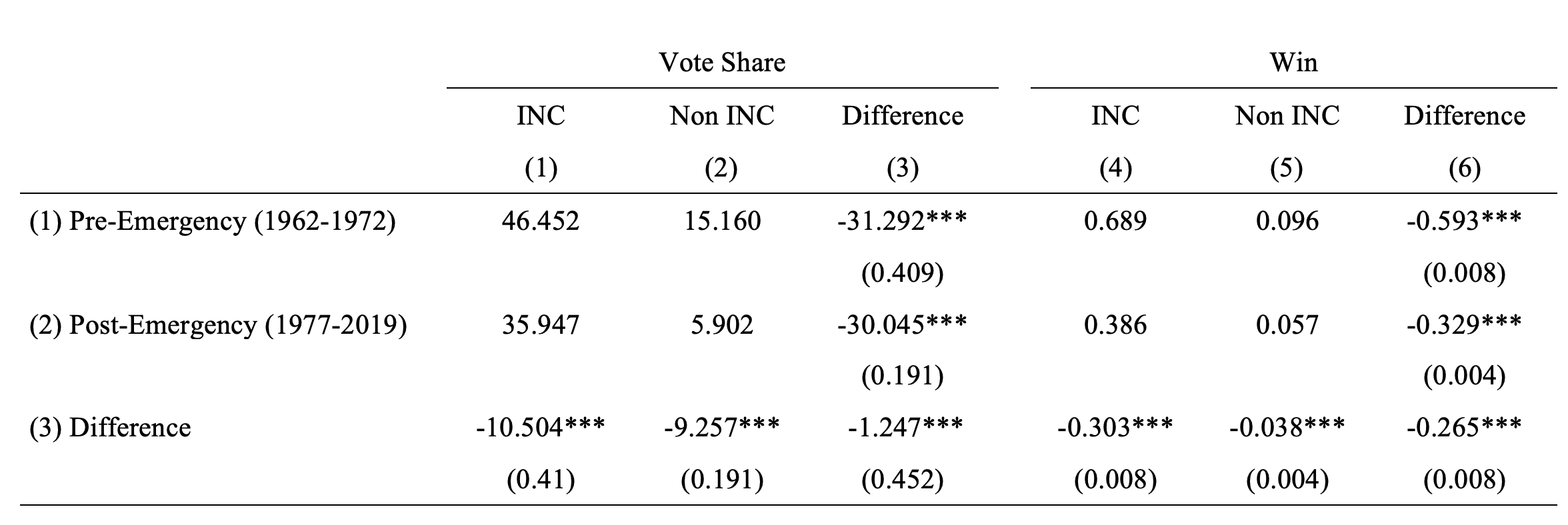}
\end{center}
{\small Notes: The data are from all general elections to the lower house of the Indian parliament (Lok Sabha) between 1962 and 2019. Standard errors are in parentheses. *** p$<0.01$, ** p$<0.05$, * p$<0.1$
}
\end{table}


\clearpage
\begin{table}[htbp]
\begin{center}
\caption{\label{figure:Table2}\textbf{Effect of Authoritarianism on Vote Share}}
\includegraphics[width=\textwidth]{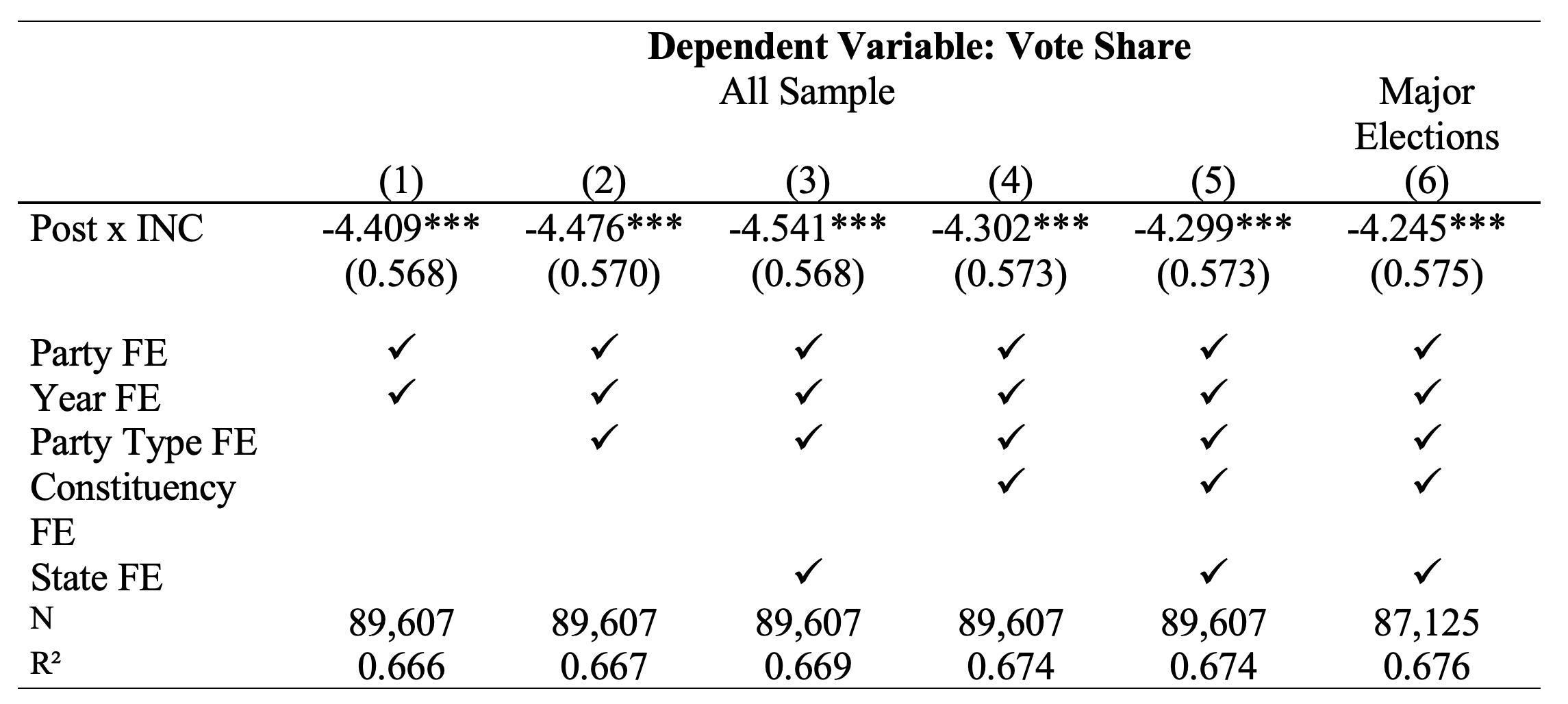}
\end{center}
{\small Notes: The data are from all general elections to the lower house of the Indian parliament (Lok Sabha) between 1962 and 2019. The unit of observation is a candidate. Each column reports estimates from OLS regression. Party FE includes the dummy variable of each party from which the candidate is contesting (independent party if the candidate has no party affiliation). Year FE includes election year fixed effects. Party Type FE includes the dummy variable of types of the political party (such as a national, state-based, local party), which proxies for geographical representation of a party. Constituency FE and State FE include the dummy variable of each constituency and state from which the candidate is contesting, respectively. Robust standard errors in parentheses clustered at constituency level. *** p$<0.01$, ** p$<0.05$, * p$<0.1$
}
\end{table}


\clearpage
\begin{table}[htbp]
\begin{center}
\caption{\label{figure:Table3}\textbf{Effect of Authoritarianism on Election Win}}
\includegraphics[width=\textwidth]{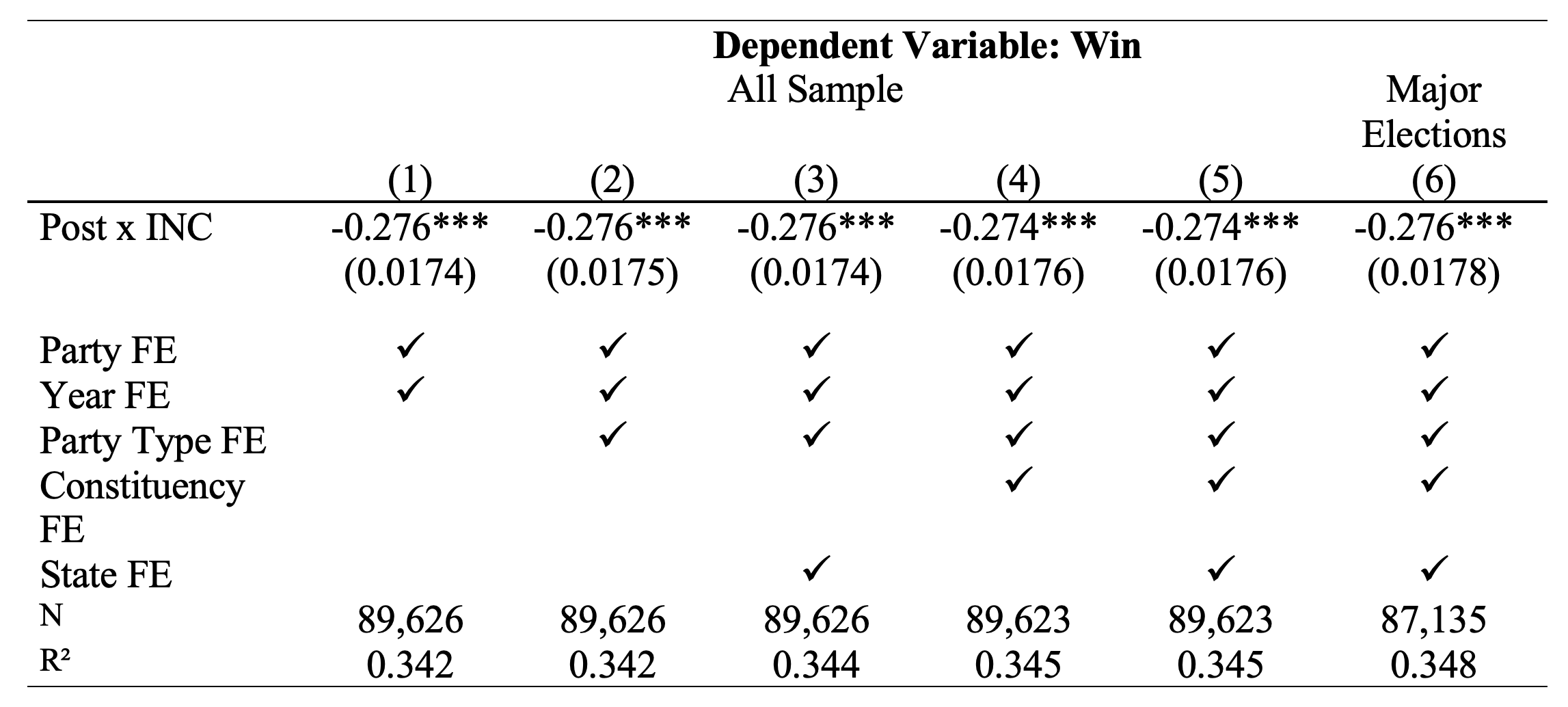}
\end{center}
{\small Notes: The data are from all general elections to the lower house of the Indian parliament (Lok Sabha) between 1962 and 2019. The unit of observation is a candidate. Each column reports estimates from OLS regression. Party FE includes the dummy variable of each party from which the candidate is contesting (independent party if the candidate has no party affiliation). Year FE includes election year fixed effects. Party Type FE includes the dummy variable of types of the political party (such as a national, state-based, local party), which proxies for geographical representation of a party. Constituency FE and State FE include the dummy variable of each constituency and state from which the candidate is contesting, respectively. Robust standard errors in parentheses clustered at constituency level. *** p$<0.01$, ** p$<0.05$, * p$<0.1$
}
\end{table}


\clearpage
\begin{table}[htbp]
\begin{center}
\caption{\label{figure:Table4}\textbf{DDD Estimation}}
\includegraphics[width=\textwidth]{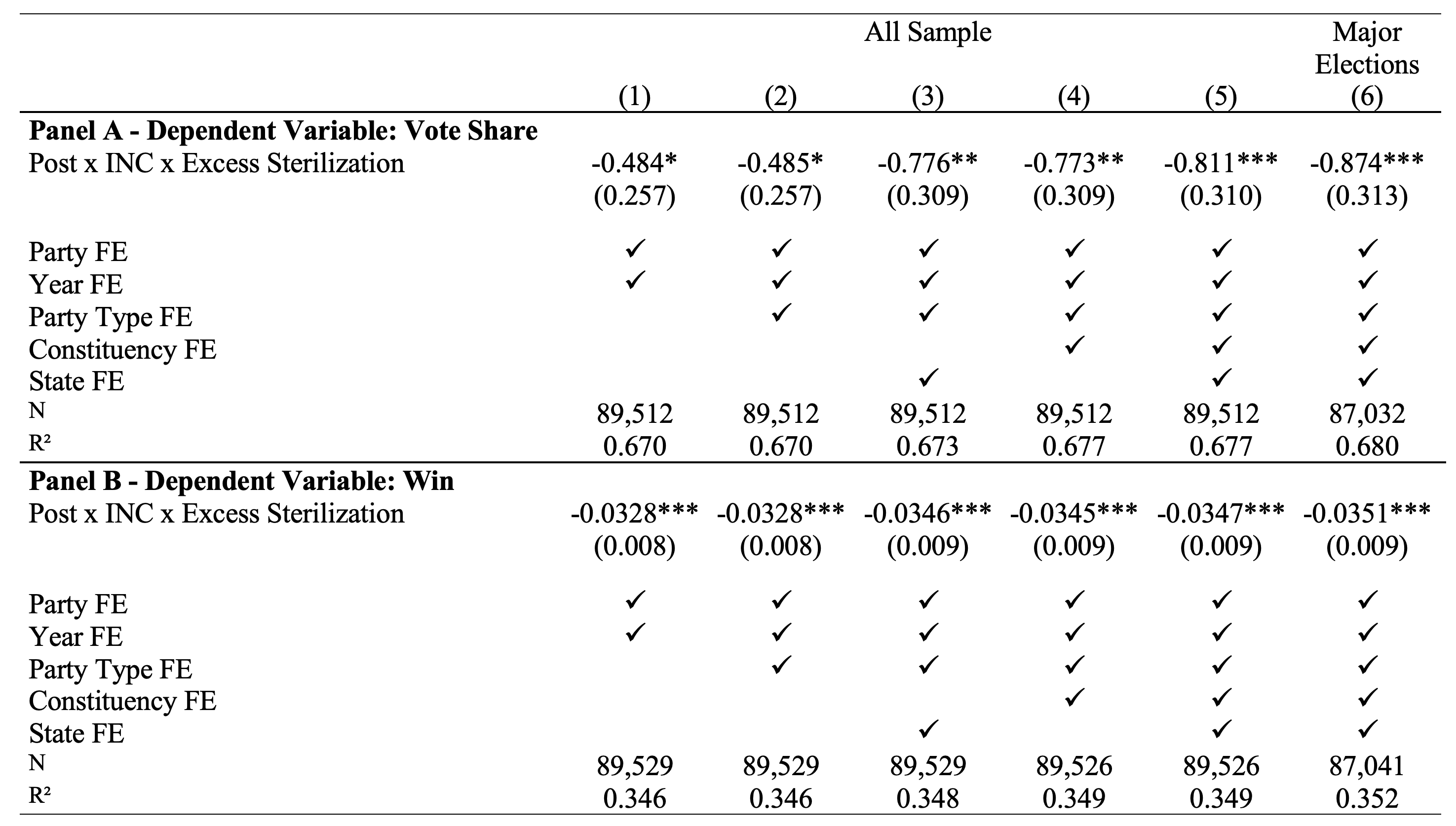}
\end{center}
{\small Notes: The data are from all general elections to the lower house of the Indian parliament (Lok Sabha) between 1962 and 2019. The unit of observation is a candidate. Each column reports estimates from OLS regression. Party FE includes the dummy variable of each party from which the candidate is contesting (independent party if the candidate has no party affiliation). Year FE includes election year fixed effects. Party Type FE includes the dummy variable of types of the political party (such as a national, state-based, local party), which proxies for geographical representation of a party. Constituency FE and State FE include the dummy variable of each constituency and state from which the candidate is contesting, respectively. Each Regression includes interaction terms to perform DDD analysis but not reported here. Robust standard errors in parentheses clustered at the constituency level. *** p$<0.01$, ** p$<0.05$, * p$<0.1$
}
\end{table}


\clearpage
\begin{sidewaystable}[htbp]
\begin{center}
\caption{\label{figure:Table5}\textbf{Effect of Authoritarianism on Voter Turnout}}
\includegraphics[width=\textwidth]{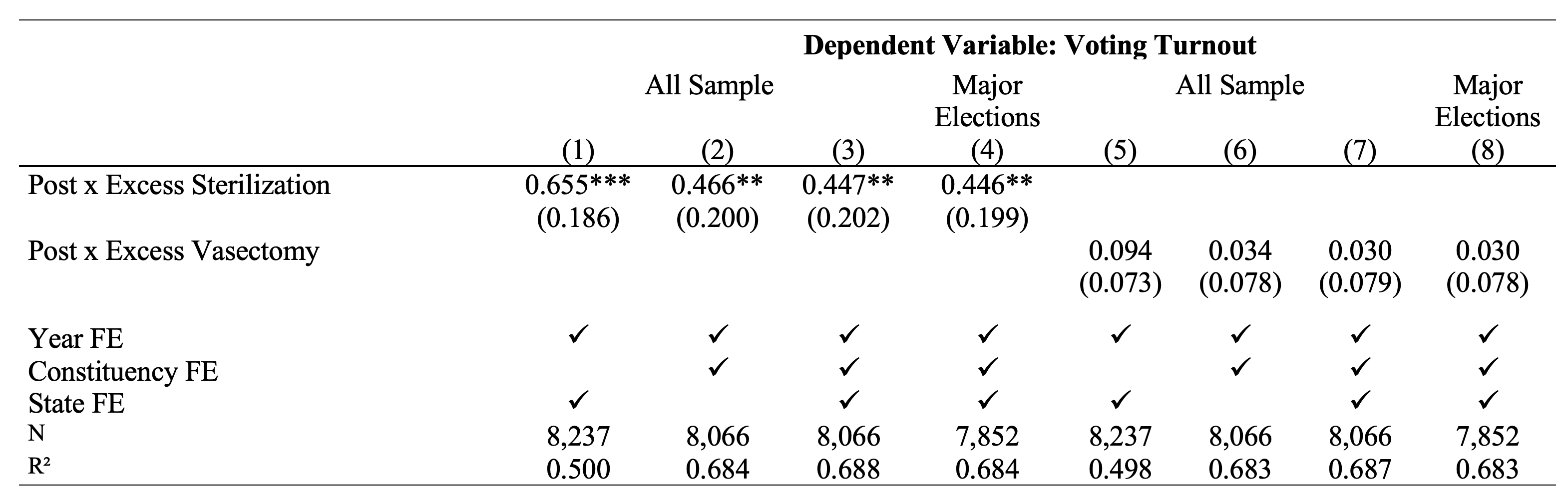}
\end{center}
{\small Notes: The data are from all general elections to the lower house of the Indian parliament (Lok Sabha) between 1962 and 2019. The unit of observation is a constituent-year. Each column reports estimates from OLS regression. Year FE includes election year fixed effects. Constituency FE and State FE include the dummy variable of each constituency and state from which the candidate is contesting, respectively. Robust standard errors in parentheses clustered at constituency level. *** p$<0.01$, ** p$<0.05$, * p$<0.1$
}
\end{sidewaystable}


\clearpage
\begin{sidewaystable}[htbp]
\begin{center}
\caption{\label{figure:Table6}\textbf{Effect of Authoritarianism on Number of Candidates Contested in Elections}}
\includegraphics[width=\textwidth]{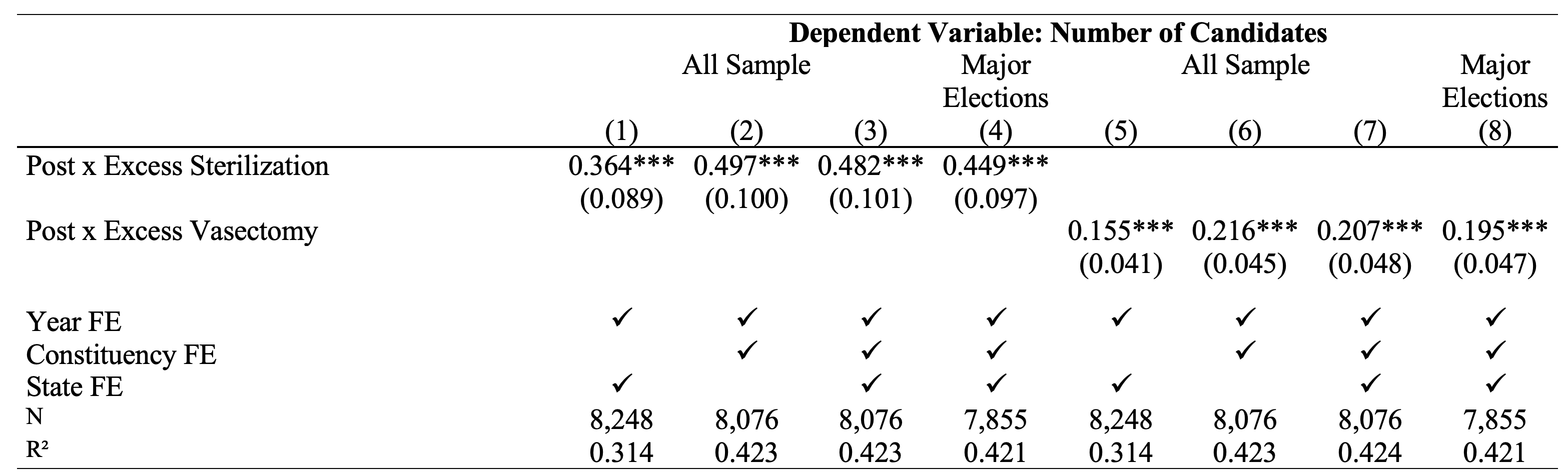}
\end{center}
{\small Notes: The data are from all general elections to the lower house of the Indian parliament (Lok Sabha) between 1962 and 2019. The unit of observation is a constituent-year. Each column reports estimates from OLS regression. Year FE includes election year fixed effects. Constituency FE and State FE include the dummy variable of each constituency and state from which the candidate is contesting, respectively. Robust standard errors in parentheses clustered at constituency level. *** p$<0.01$, ** p$<0.05$, * p$<0.1$
}
\end{sidewaystable}


\clearpage
\begin{table}[htbp]
\begin{center}
\caption{\label{figure:Table7}\textbf{Effect of Authoritarianism on Confidence in Politicians}}
\includegraphics[width=0.7 \textwidth]{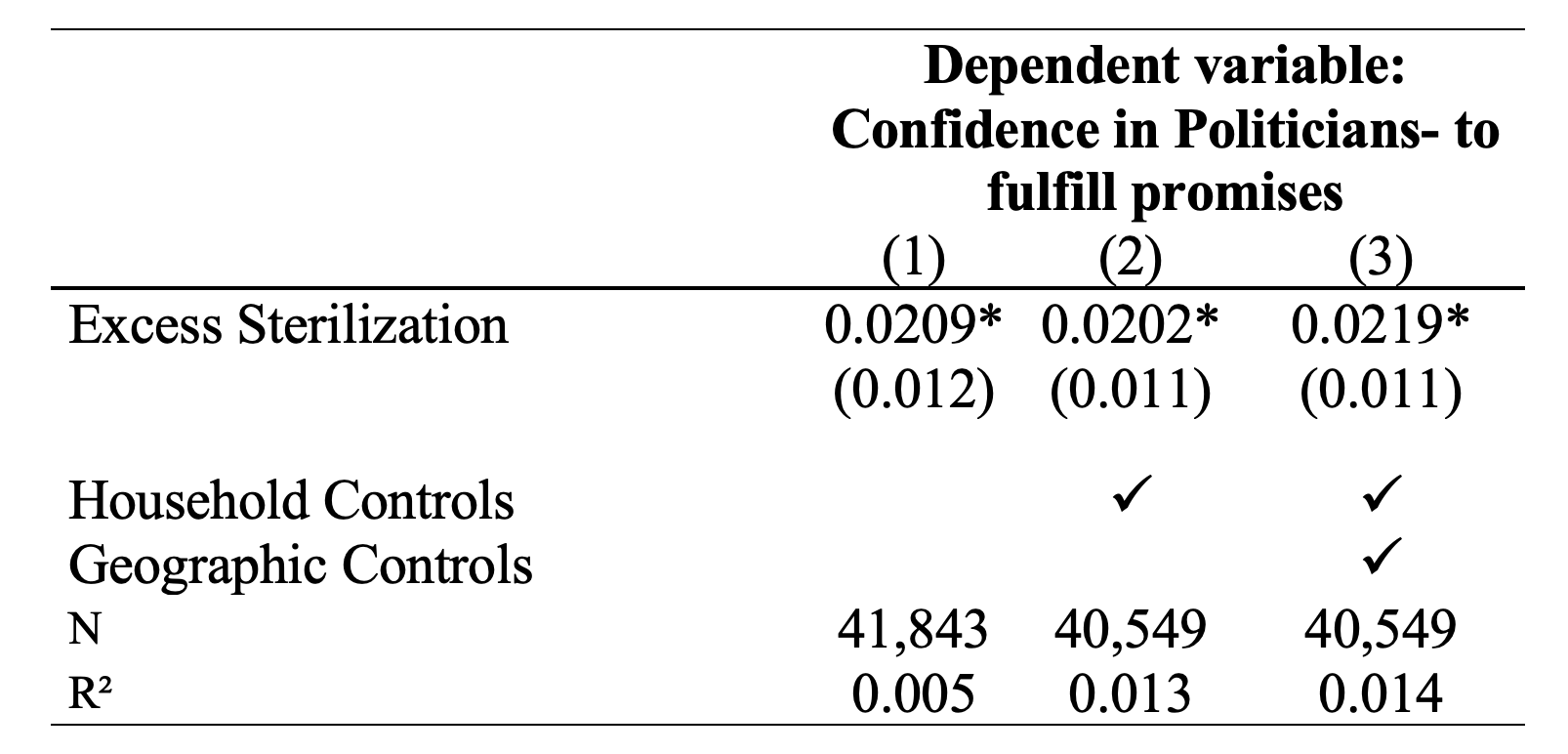}
\end{center}
{\small Notes: The data are from IHDS-II. The unit of observation is a household. The IDHS-II assigns the value 1 to “a great deal of confidence,” 2 to “only some confidence,” and 3 to “hardly any confidence at all.” Therefore, a higher score constitutes a lower level of confidence. Each column reports estimates from OLS regression. The household controls include household size, income, ten sources of main income fixed effects, eight religion fixed effects, five caste fixed effects, two wealth class fixed effects (poor, middle class, (comfortable as the reference group)), 16 education of the household head fixed effects, an indicator for whether any household member is covered by government health insurance, and an indicator for whether the household has a BPL card. The geographic controls include state-level population density (in log) and three places of residence fixed effects. Robust standard errors in parentheses clustered at the state level. *** p$<0.01$, ** p$<0.05$, * p$<0.1$
}
\end{table}

\appendix
\renewcommand\thefigure{\thesection.\arabic{figure}}
\renewcommand\thetable{\thesection.\arabic{table}}
\setcounter{figure}{0}
\setcounter{table}{0}

\clearpage
\pagenumbering{roman}

\title{\textbf{ \Large {Online Appendix \\
\vskip 1cm
  The Legacy of Authoritarianism in a Democracy
}}}

\clearpage
\section{APPENDIX Figures}

\begin{figure}[htbp]
\begin{center}
\caption{\label{figure:FigureA1}\textbf{Countries Becoming Substantially Autocratic and Democratic (2010--2020) }}

\includegraphics[width=\textwidth]{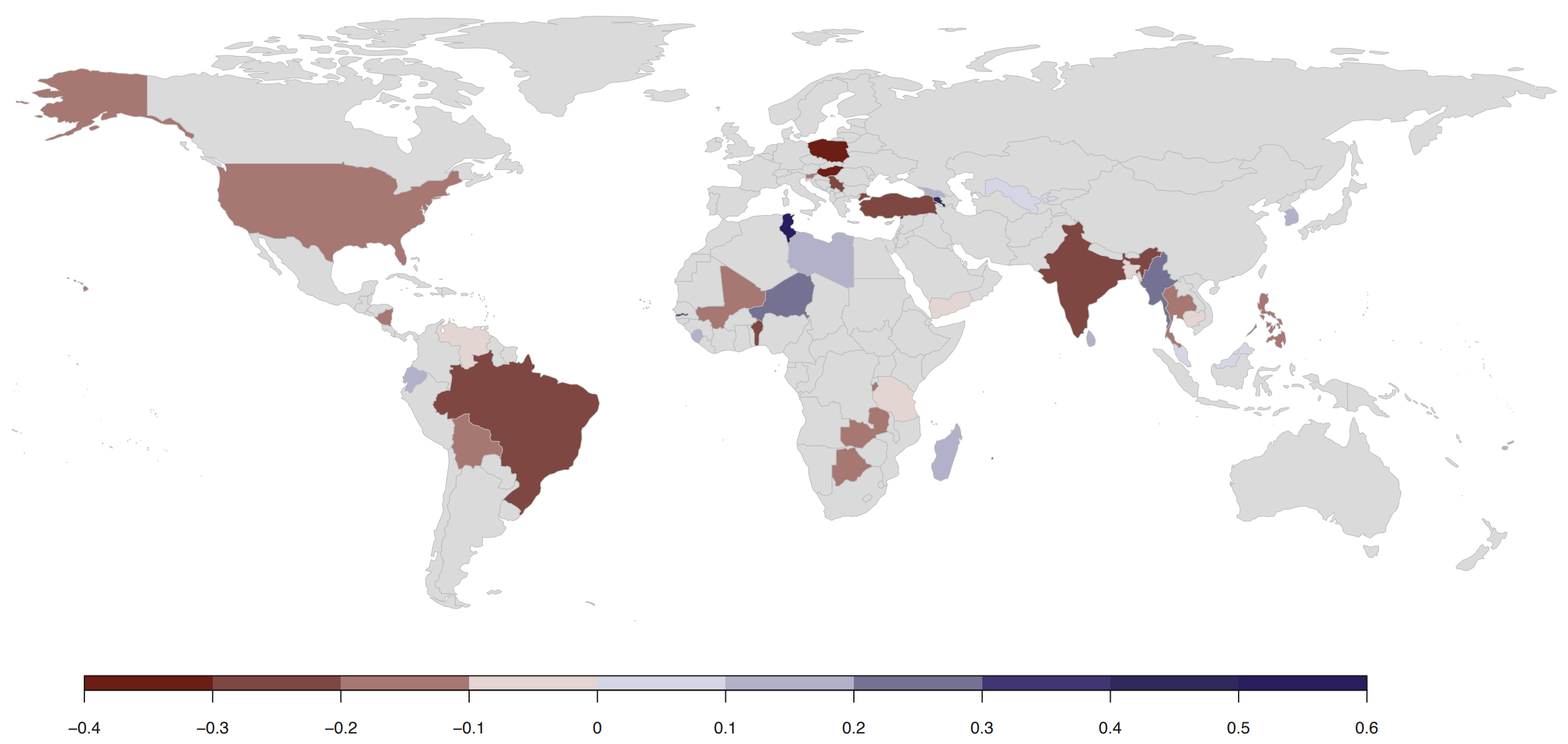}
\end{center}
{\small Note: The figure presents the countries that have become substantially autocratic or democratic between 2010 and 2020. Darker red indicates countries where the Liberal Democracy Index (LDI) has declined substantially and significantly over 2010 and 2020. Darker blue indicates countries where the level of democracy has advanced substantially during this period. Countries in gray are substantially unchanged. \\
Data Source: Varieties of Democracy (V-Dem). https://www.v-dem.net/en/ Accessed on 20th July 2021

}
\end{figure}

\clearpage
\begin{figure}[htbp]
\begin{center}
\caption{\label{figure:FigureA2}\textbf{Total Number of Sterilizations and Types of Sterilizations Performed in India (1956-82)}}
\includegraphics[width=\textwidth]{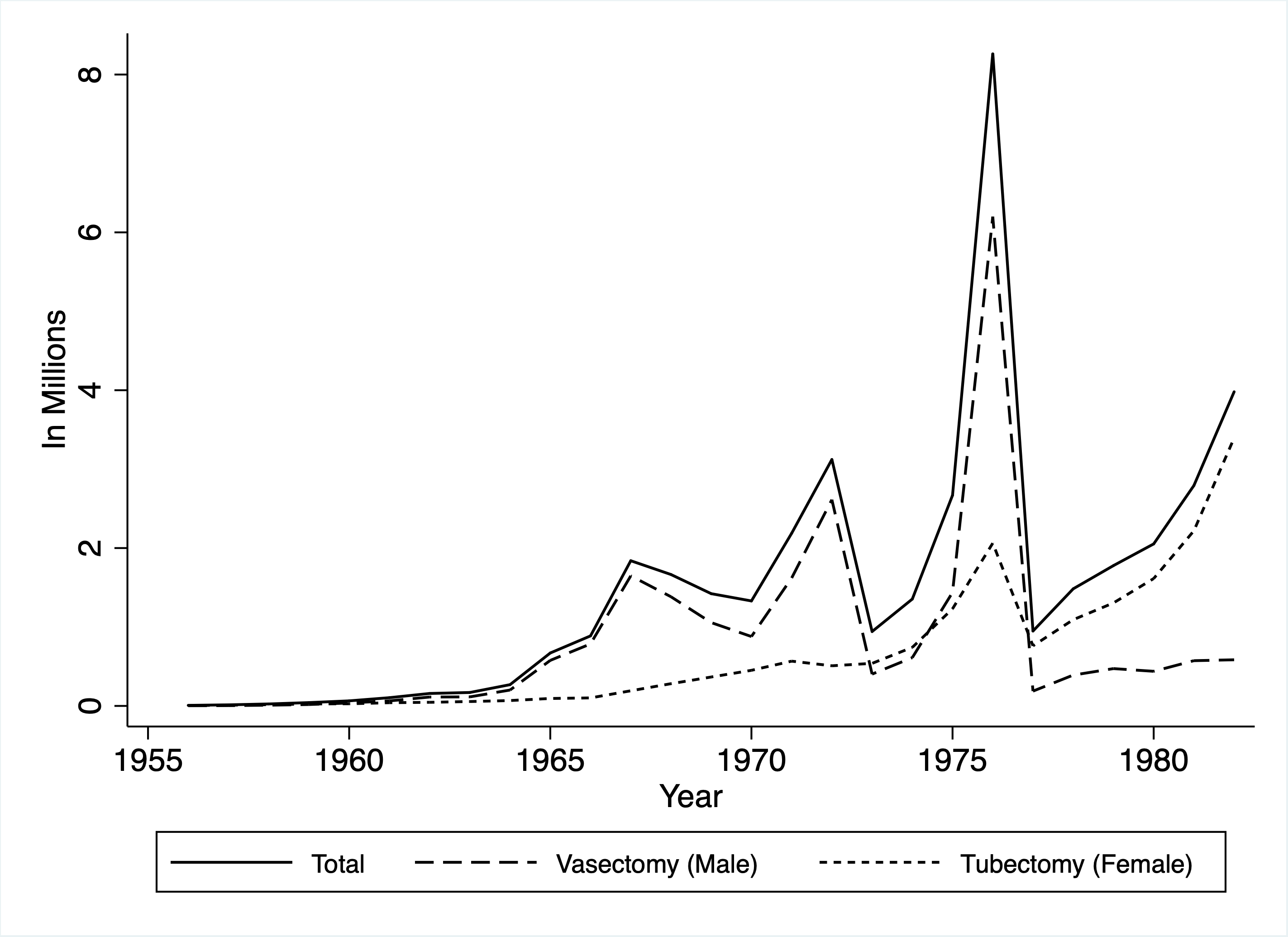}
\end{center}
{\small Notes: The figure presents the total number of sterilizations along with the types of sterilization performed in India every year since the beginning of the program in 1956. The solid line represents the total number of sterilizations performed every year. The dashed and short dashed lines represent the total number of vasectomies and tubectomies performed every year, respectively.

}
\end{figure}


\clearpage
\begin{figure}[htbp]
\begin{center}
\caption{\label{figure:FigureA3}\textbf{The Effects of Authoritarianism on Voting Behavior (Major Elections Only) }}
\subcaption{Panel A: The Effects of Authoritarianism on Vote Share}
\includegraphics[height=8cm]{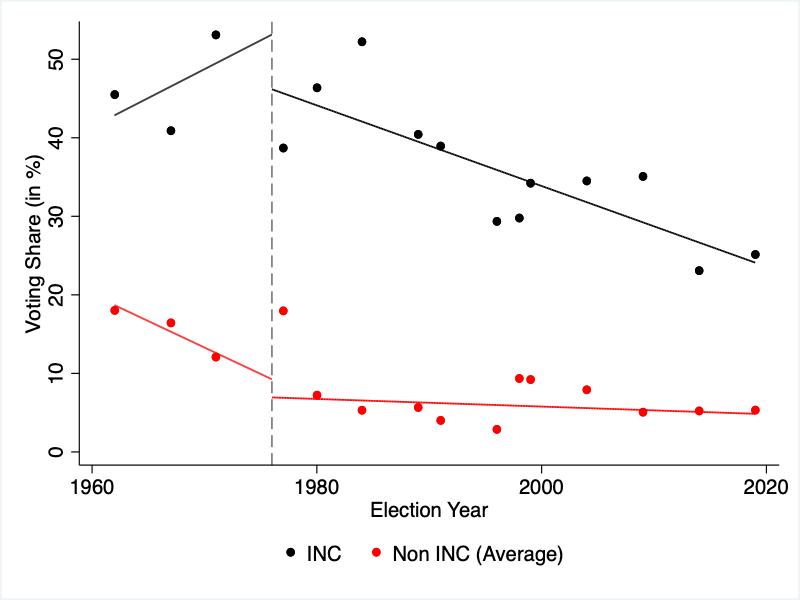}
\subcaption{Panel B: The Effects of Authoritarianism on Probability of Win}
\includegraphics[height=8cm]{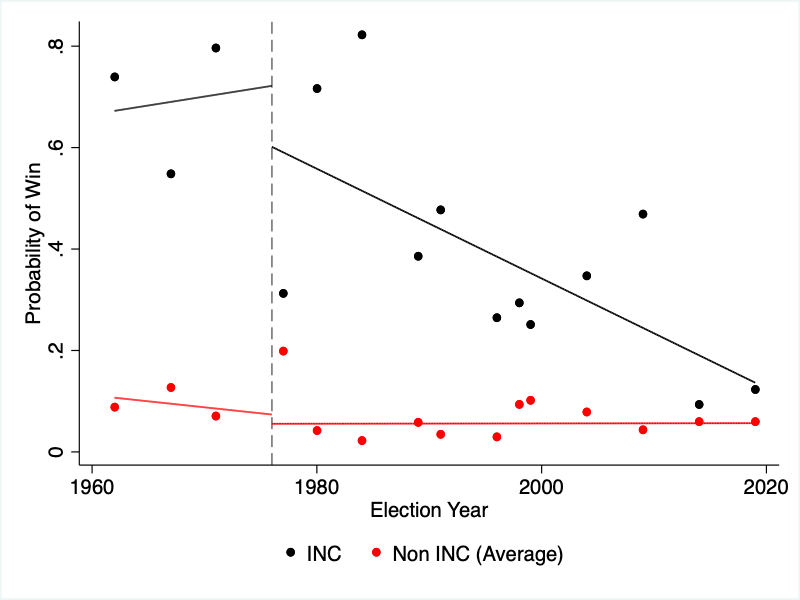}
\end{center}
{\small Notes: The figure plots the effects of authoritarianism on voting behavior and election outcomes of major elections in India. The data are from all major general elections to the lower house of the Indian parliament (Lok Sabha) between 1962 and 2019. Panel A plots the average vote share of INC and non-INC candidates received in a constituency in each major election year. Panel B plots the probability of which INC and non-INC candidates won in a constituency in each major election year. The dashed line represents the end of the authoritarian rule in India. The average of INC and non-INC voting share and probability of win does not sum to 100 and 1 respectively, because there are more than one non-INC candidates contesting for election in a constituency.
}
\end{figure}


\clearpage
\begin{figure}[htbp]
\begin{center}
\caption{\label{figure:FigureA4}\textbf{DDD Estimates: Event Study (All Election Years) }}
\subcaption{Panel A: Vote Share}
\includegraphics[height=8cm]{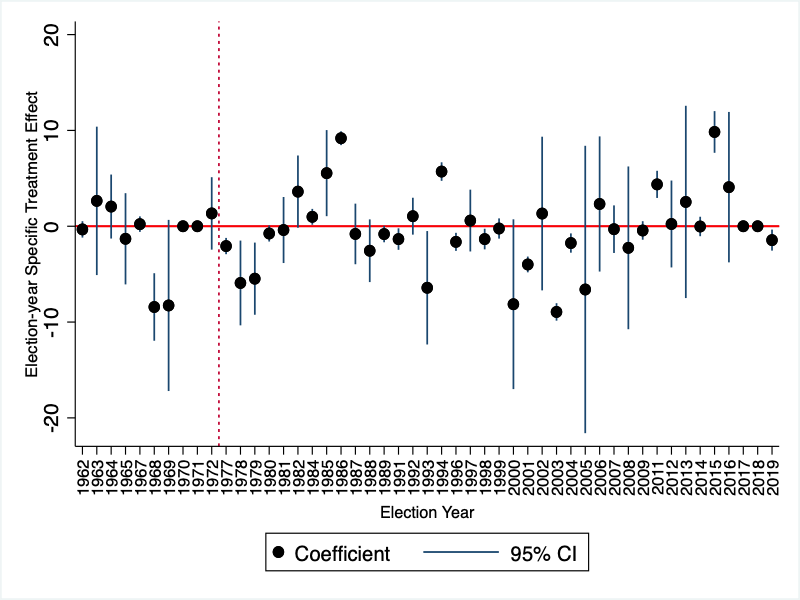}
\subcaption{Panel B: Probability of Win}
\includegraphics[height=8cm]{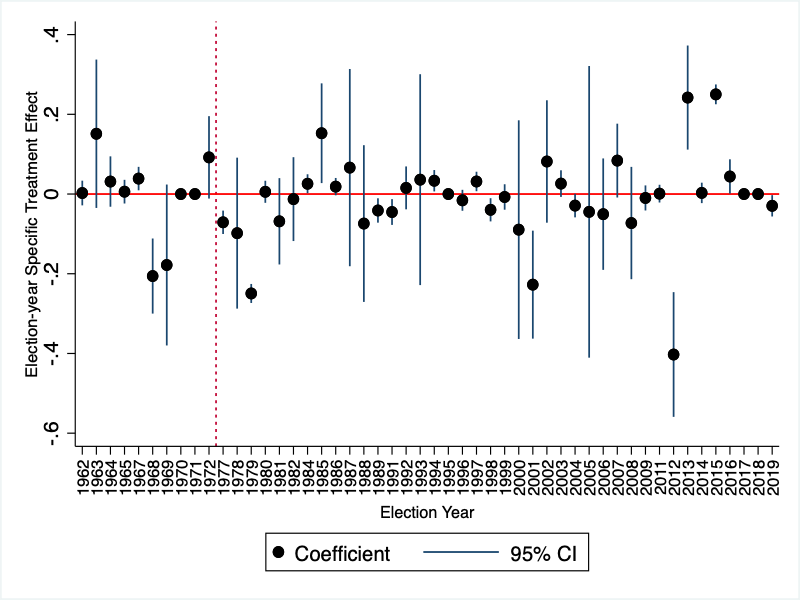}
\end{center}
{\small Notes: The figure presents the political consequences of authoritarianism in India through an event study framework. The data are from all general elections to the lower house of the Indian parliament (Lok Sabha) between 1962 and 2019. Panel A plots the year-specific treatment effects of INC candidates' vote share. Panel B presents the year-specific treatment effects of INC candidates' probability of winning an election. The (red) dashed line represents the end of the authoritarian rule in India.
}
\end{figure}


\clearpage
\begin{sidewaysfigure}[htbp]
\begin{center}
\caption{\label{figure:FigureA5}\textbf {Robustness to RDD Estimates: Using Month as Running Variable}}
\includegraphics[height=13cm]{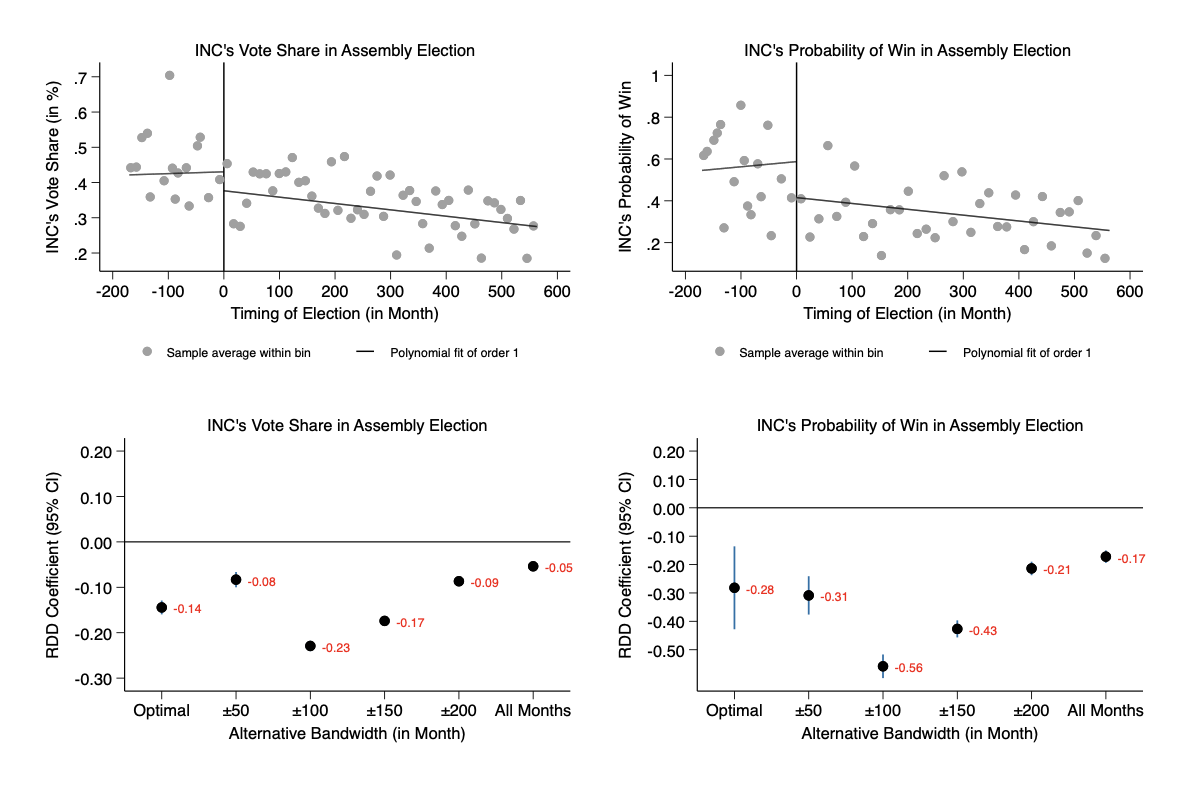}
\end{center}
{\small Notes: The figure presents RDD estimates. The data are from the state-level Assembly Elections between 1962 and 2023. The running variable is the election month. The left panel plots the RDD results of INC's vote share with alternative bandwidths in months. The right panel plots the RDD results of the INC candidate's probability of winning with alternative bandwidths in months. I set the local polynomial of order 1, which is simple, transparent, and easy to interpret (using a local polynomial of order 2 also produces similar results). I use a uniform kernel (using a triangular kernel also produces identical results). 
}
\end{sidewaysfigure}


\clearpage
\begin{figure}[htbp]
\begin{center}
\caption{\label{figure:FigureA6}\textbf{DID Estimates: Event Study (INC Candidates Only) }}
\subcaption{Panel A: Vote Share}
\includegraphics[height=8cm]{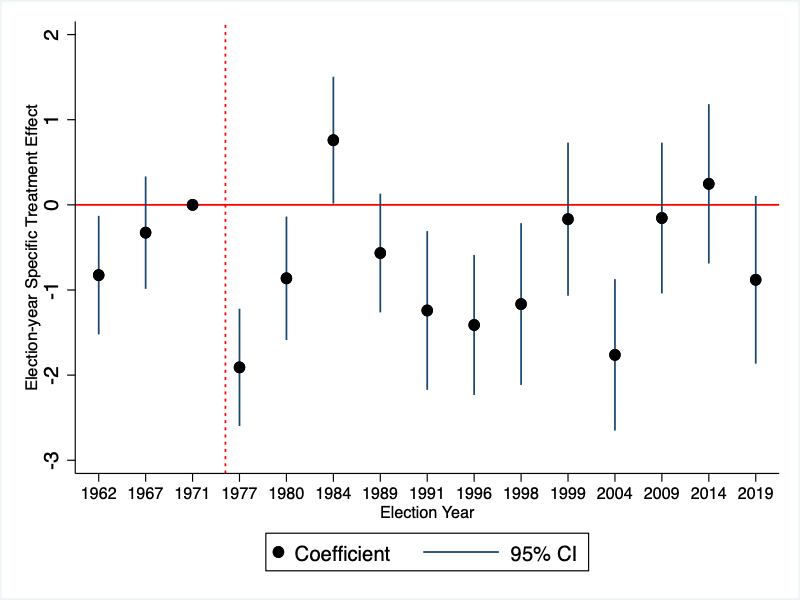}
\subcaption{Panel B: Probability of Win}
\includegraphics[height=8cm]{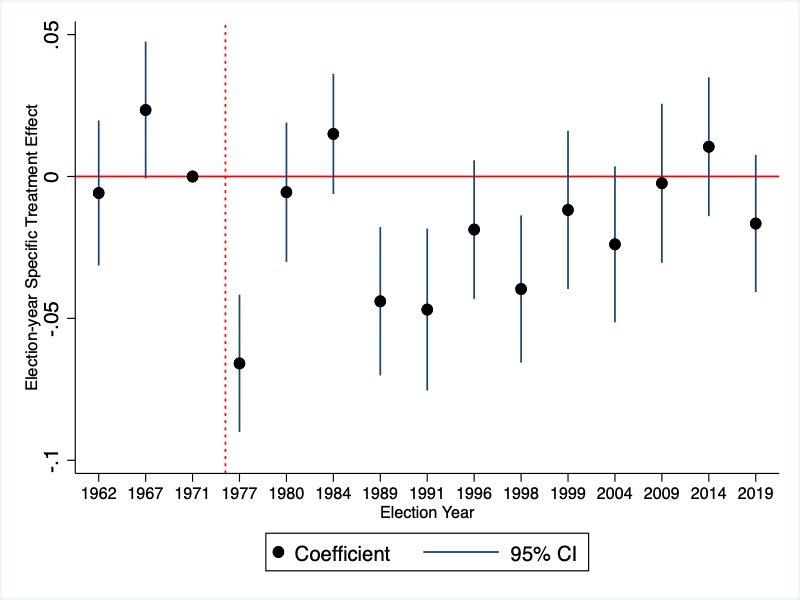}
\end{center}
{\small Notes: The figure presents the political consequences of authoritarianism in India through an event study framework restricting to INC candidates only (considering years interacted with excess sterilization). The election data are from all major general elections to the lower house of the Indian parliament (Lok Sabha) between 1962 and 2019. Panel A plots the year-specific treatment effects of vote share. Panel B presents the year-specific treatment effects of the probability of winning an election. The (red) dashed line represents the end of the authoritarian rule in India.
}
\end{figure}

\clearpage
\begin{figure}[htbp]
\begin{center}
\caption{\label{figure:FigureA7}\textbf{Change in Democracy Index in Brazil, Hungary, the Philippines, Poland, and Turkey}}

\includegraphics[width=\textwidth]{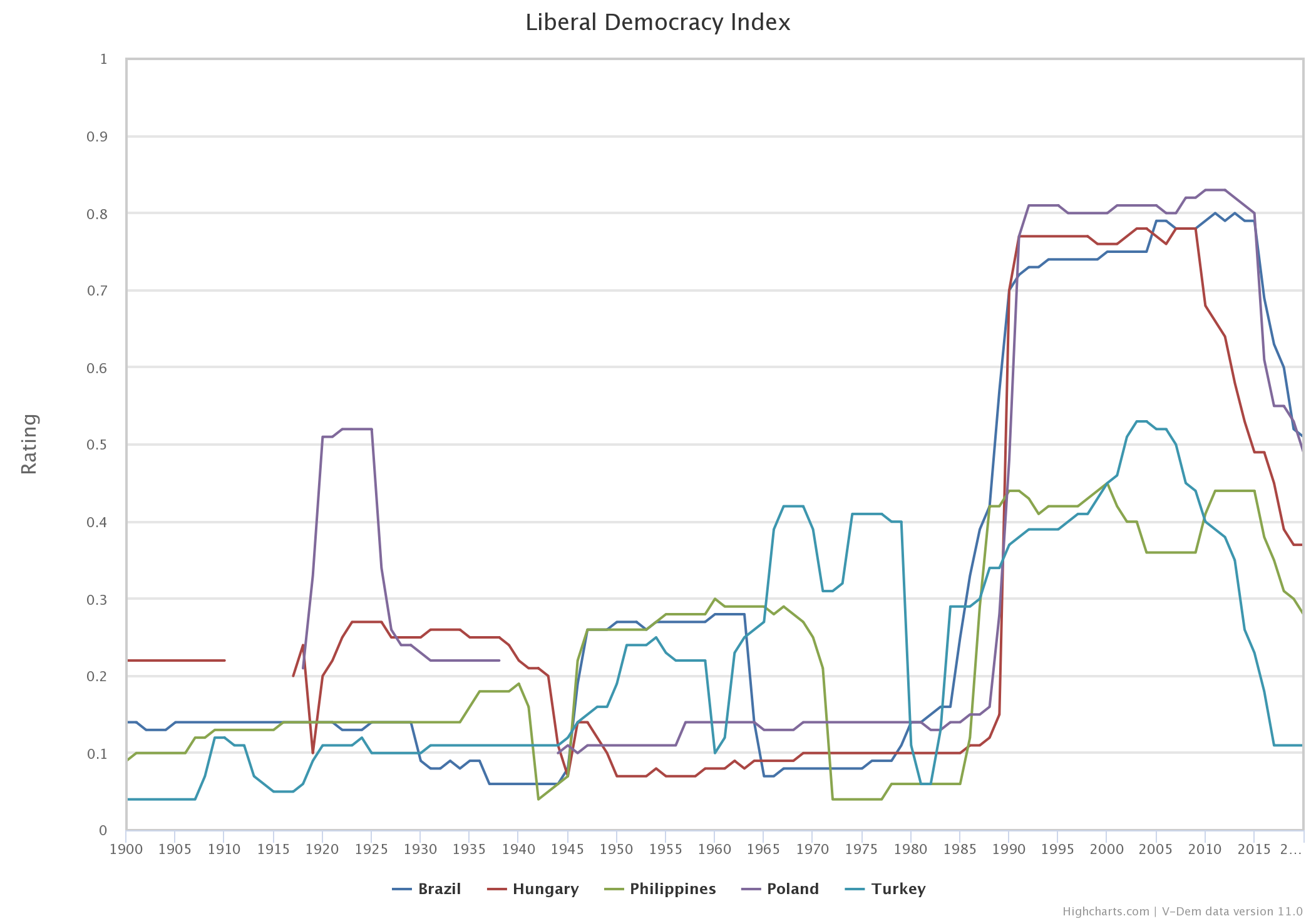}
\end{center}
{\small Notes: The figure presents the change in the Liberal Democracy Index in Brazil, Hungary, the Philippines, Poland, and Turkey between 1900 and 2020. \\
Data Source: Varieties of Democracy (V-Dem). https://www.v-dem.net/en/ Accessed on 20th July 2021
}
\end{figure}

\clearpage
\section{APPENDIX Tables}
\begin{table}[htbp]
\begin{center}
\caption{\label{figure:TableB1}\textbf{Flexible DDD Approach (Tabular Form)}}
\includegraphics[width=0.9 \textwidth]{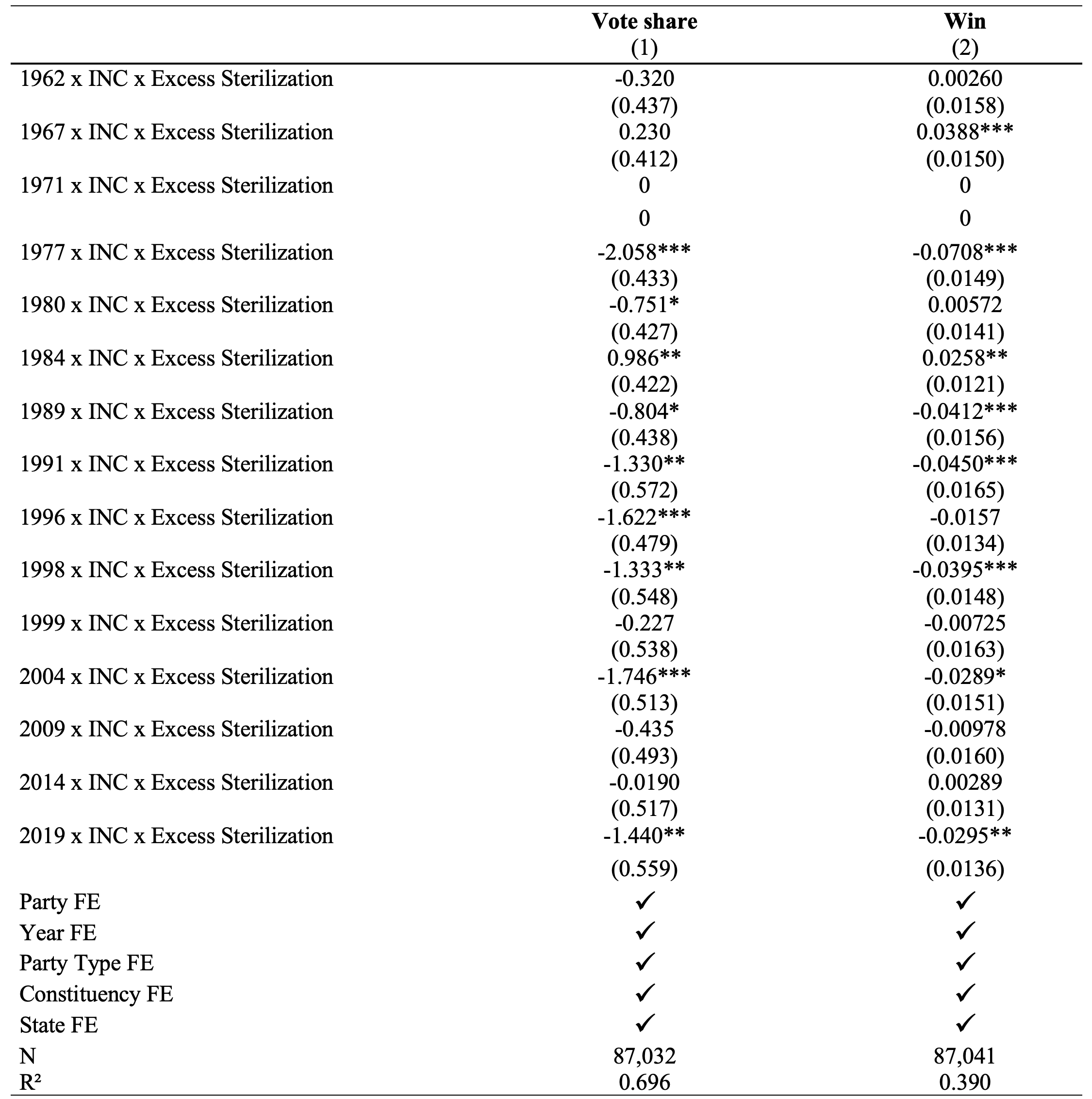}
\end{center}
{\small Notes: The data are from all major general elections to the lower house of the Indian parliament (Lok Sabha) between 1962 and 2019. The unit of observation is a candidate. Each column reports estimates from OLS regression. Party FE includes the dummy variable of each party from which the candidate is contesting (independent party if the candidate has no party affiliation). Year FE includes election year fixed effects. Party Type FE includes the dummy variable of types of the political party (such as a national, state-based, local party), which proxies for geographical representation of a party. Constituency FE and State FE include the dummy variable of each constituency and state from which the candidate is contesting, respectively. Each Regression includes interaction terms to perform DDD analysis but not reported here. Robust standard errors in parentheses clustered at constituency level. *** p$<0.01$, ** p$<0.05$, * p$<0.1$
}
\end{table}

\clearpage
\begin{table}[htbp]
\begin{center}
\caption{\label{figure:TableB2}\textbf{Robustness to DDD Estimates Using Alternative Measures of Force Sterilization Policy - Male Sterilization}}
\includegraphics[width=\textwidth]{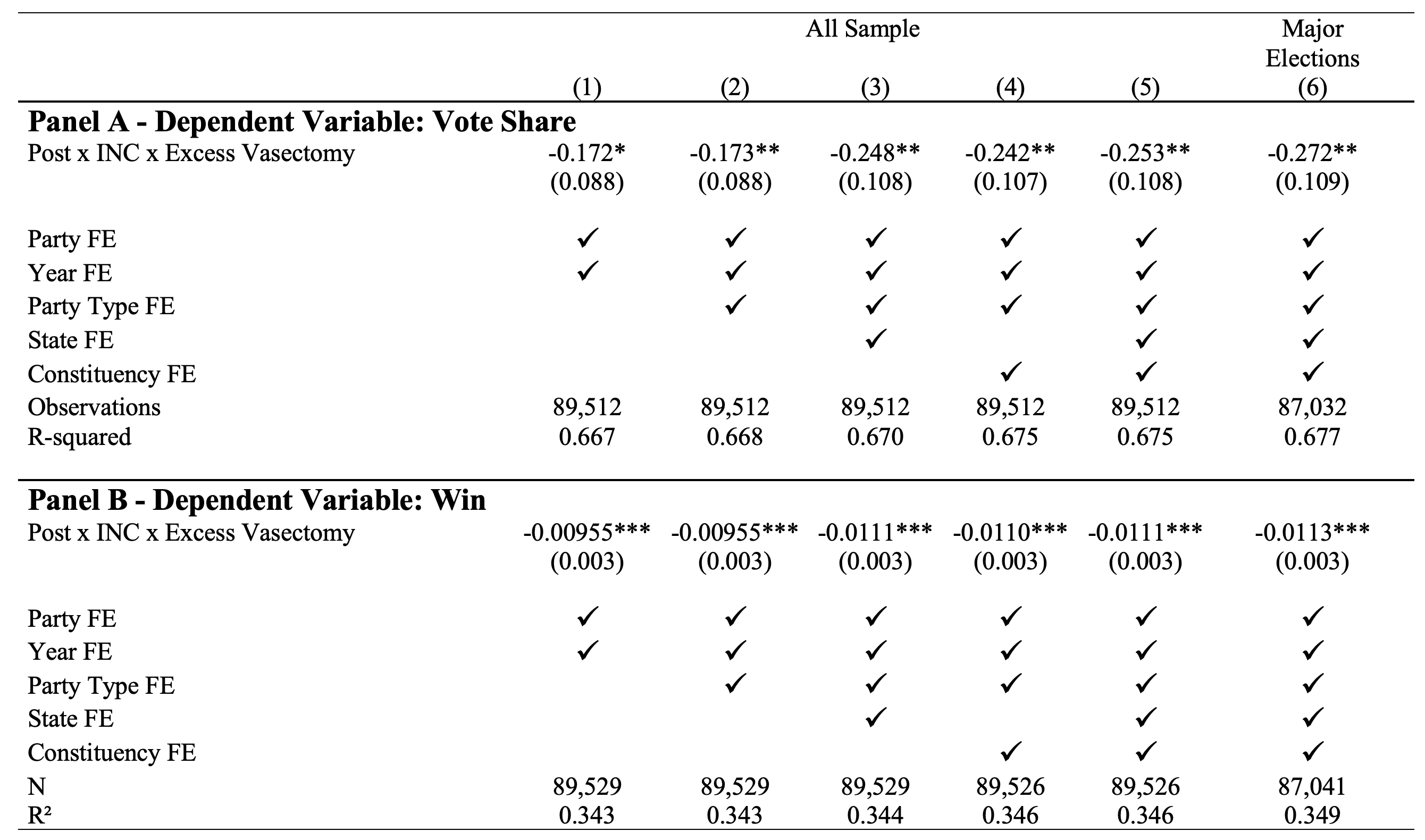}
\end{center}
{\small Notes: The data are from all general elections to the lower house of the Indian parliament (Lok Sabha) between 1962 and 2019. The unit of observation is a candidate. Each column reports estimates from OLS regression. Party FE includes the dummy variable of each party from which the candidate is contesting (independent party if the candidate has no party affiliation). Year FE includes election year fixed effects. Party Type FE includes the dummy variable of types of the political party (such as a national, state-based, local party), which proxies for geographical representation of a party. Constituency FE and State FE include the dummy variable of each constituency and state from which the candidate is contesting, respectively.  Each Regression includes interaction terms to perform DDD analysis but not reported here. Robust standard errors in parentheses clustered at constituency level. *** p$<0.01$, ** p$<0.05$, * p$<0.1$
}
\end{table}

\end{document}